\documentclass[reqno]{amsart}

\usepackage[T1]{fontenc}
\usepackage{xpatch}

\patchcmd{\section}{\normalfont}{\normalfont\LARGE\scshape}{}{}
\patchcmd{\subsection}{\normalfont}{\alessandrofont}{}{}
\patchcmd{\subsubsection}{\normalfont}{\alessandrofont}{}{}
\newcommand{\alessandrofont}{\large\sffamily}

\usepackage{newtxtext}
\usepackage{amssymb}
\usepackage{amsmath}

\usepackage{natbib}
\usepackage{graphicx}

\usepackage[tikz]{bclogo}
\presetkeys{bclogo}{
ombre=true,
epBord=3,
couleur = blue!15!white,
couleurBord = red,
arrondi = 0.2,
logo=\bctrombone
}{}

\numberwithin{equation}{section} \oddsidemargin=-.0cm
\usepackage{xcolor,colortbl}
\usepackage[colorlinks=true, pdfstartview=FitV, linkcolor=blue,
            citecolor=blue, urlcolor=blue]{hyperref}  % the hyperref package should be the last package in order the link to 	
\usepackage{color,amsthm}

\usepackage{epsfig}
\usepackage[mathscr]{eucal}
\usepackage{lipsum}

\newcommand{\bk}{\ensuremath {\boldsymbol {k}}}

\newcommand{\bS}{\ensuremath {\boldsymbol {S}}}
\newcommand{\bV}{\ensuremath {\boldsymbol {V}}}
\newcommand{\bv}{\ensuremath {\boldsymbol {v}}}

\newcommand{\bI}{\ensuremath {\boldsymbol {\mathcal{I}}}}

\def\dd{{\rm d}}

\def\beq{\begin{equation}}
\def\eeq{\end{equation}}

\newcommand{\E}{\mathscr{E}}

\renewcommand{\L}{\mathsf{L}}

\newcommand{\Z}{\mathscr{Z}}

\def\bi{\begin{itemize}}
\def\ei{\end{itemize}}

\definecolor{rred}{rgb}{0.7,0,0.1}
\newcommand{\mk}{\color{black}}
\newcommand{\mkr}{\color{black}}%%%magenta

\definecolor{greenrb}{rgb}{0.2,0.6,0.2}
\begin{document}

%%%%%%%%%%%%%%%%%%%%%%%%%%%%%%%%%%%%%%%%%%%%%%%
\title[Turbulence closure with small, local neural networks]{Turbulence closure with small, local neural networks: Forced two-dimensional and $\beta$-plane flows}

%%%%%%%%%%%%%%%%%%%%%%%%%%%%%%%%%%%%%%%%%%%%%%%%%%%%%%
\author{Kaushik Srinivasan}
\address{(KS) Department of Atmospheric and Oceanic Sciences, University of California, Los Angeles, CA 90095-1565, USA}
\email{kaushiks@g.ucla.edu}

%%%%%%%%%%%%%%%%%%%%%%%%%%%%%%%%%%%%%%%%%%%%%%%%%%%%%%

\author{Micka{\"e}l D. Chekroun}
\address{(MDC) Department of Atmospheric and Oceanic Sciences, University of California, Los Angeles, CA 90095-1565, USA, and Department of Earth and Planetary Sciences, Weizmann Institute, Rehovot 76100, Israel}

%%%%%%%%%%%%%%%%%%%%%%%%%%%%%%%%%%%%%%%%%%%%%%%%%%%%%%

\author{James C. McWilliams}
\address{(JCM) Department of Atmospheric and Oceanic Sciences and Institute of Geophysics and Planetary Physics, University of California, Los Angeles, CA 90095-1565, USA} 

\begin{abstract}
{\small
	We parameterize sub-grid scale (SGS) fluxes in sinusoidally forced
	two-dimensional turbulence on the $\beta$-plane at high Reynolds numbers (Re$\sim$25000)
	using simple 2-layer Convolutional Neural Networks (CNN) having only O(1000)
	parameters, two orders of magnitude smaller than recent studies employing
	deeper CNNs with 8-10 layers; we obtain stable, accurate, and long-term online
	or \textit{a posteriori} solutions at 16X downscaling factors. Our methodology
	significantly improves training efficiency and speed of online Large Eddy
	Simulations (LES) runs, while offering insights into the physics of closure in
	such turbulent flows. Our approach benefits from extensive hyperparameter
	searching in learning rate and weight decay coefficient space, as well as the
	use of cyclical learning rate annealing, which leads to more robust and
	accurate online solutions compared to fixed learning rates. Our CNNs use either
	the coarse velocity or the vorticity and strain fields as inputs, and output
	the two components of the deviatoric stress tensor, $\bS_d$. We minimize a loss
	between the SGS vorticity flux divergence (computed from the high-resolution
	solver) and that obtained from the CNN-modeled $\bS_d$, without requiring energy or
	enstrophy preserving constraints. The success of shallow CNNs in accurately
	parameterizing this class of turbulent flows implies that the SGS stresses have
	a weak non-local dependence on coarse fields; it also aligns with our physical
	conception that small-scales are locally controlled by larger scales such as
	vortices and their strained filaments. Furthermore, 2-layer
	CNN-parameterizations are more likely to be interpretable and generalizable
	because of their intrinsic low dimensionality.}
\end{abstract}
\maketitle

\tableofcontents

\section{Introduction}
\label{sec:intro}
Turbulent flows in physical systems span a vast range of spatial and temporal
scales; in the Earth's oceans, for example, mesoscale eddies, the dominant
reservoir of kinetic energy in the ocean, have scales of O(100 km) whereas,
small-scale three dimensional motions at the air-sea interface driven by a
combination of surface heating/cooling, wind action and the Earth's rotation
have scales of O(1m). This vast range of spatial scales is well beyond the
range of current numerical ocean models to solve given even the largest
available compute facilities. Climate models face even greater challenges
because they need to be run for years or decades to study climatic changes over
long temporal horizons. In practice, these models are run at a certain
resolution limited by available computational resources while unresolved
turbulent processes are effectively represented or \textit{parameterized}.  The
turbulent motions that are below the resolution of numerical models and need
parameterization or representation are commonly referred to as subgrid-scale
(SGS) motions, with their effect being expressed as effective fluxes (SGS
fluxes) that can be added to the equations of motion already solved by
numerical models.

Historically, parameterization of unresolved turbulent flows are computed
through a combination of empirical data, physics-based modeling and
simplified algebraic models computed using high resolution process models run
over short durations of time. For example, three-dimensional small-scale
turbulence near ocean surface is parameterized using the K-profile
parameterization \citep{larg94} framework which is a \textit{ad hoc} combination
of empirical and physics-based approaches that model various turbulent process
found at the air-sea interface, including surface convection and boundary layer
rotating-stratified shear turbulence.  Parameterizations are also found through
numerical process studies (essentially idealized numerical simulations of a
specific phenomenon), typically high-resolution Large Eddy
Simulations (LES) to fit simplified algebraic models based on combination of
dimensional analysis and physical models \citep{souza2020}.

An alternative approach spurred on by the development of machine learning methods, in
particular neural network based Deep Learning methods, along with the availability of
large amount of data from numerical simulations, is to use high
resolution process studies to generate SGS fluxes that can be accurately
modeled without recourse to simplified algebraic models. With the availability
of frameworks that allow incorporating neural network (NN) models trained in
the Python programming language into Fortran \citep{ott2020}, the language of
most weather, ocean and climate models, one can in principle accurately model
SGS fluxes in climate models, substantially adding to their predicability and
fidelity and reducing their biases. In practice, a major issue that manifests
is that the numerical models that incorporate NN-based models are subjected to
challenges in numerical stability \citep{brenowitz2020}.

Neural networks are hierarchical non-linear functions of simpler modular units
containing a large number of tunable-parameters that can in principle
represent any arbitrary non-linear function given sufficient data
\citep{hornik1989, zhou2020}.  Typically complex models like these can be subject to
over-fitting to a given dataset and not learn general relationships that are
actually present in the data. However, a multitude of regularization techniques
have been discovered over the past decade that allow NNs to learn non-linear
relationships that generalize well across unseen data \citep{srivastava2014,
ioffe2015, ba2016, loshchilov2016}. Applications of NNs to dynamical modeling,
in particular to the solution of complex dynamical equations that govern the
climate system, face the additional requirements that solutions of equations be
numerically stable and also that they not accrue unphysical biases over long
time horizons. 

NN-based parameterizations in dynamical systems can classified based on how the
learning framework is designed. Online or \textit{a posteriori} learning
\citep{ma2018, rasp2020} directly incorporates the NN-parameterization into the
equations of motion; the equations of motions are time-stepped forward and the
parameters of the NN are learnt subject to the constraint that NN-driven flow
trajectory matches the outputs of a high resolution numerical solution
truncated to the resolution of the NN-based solutions. This approach, a form of
\textit{trajectory optimization} that is common in imitation learning
\citep{hussein2017} and has the advantage that the NN-driven solution is
physics-aware, promises that well optimized trajectories are not subject to
biases not present in the high resolution numerical solution. A drawback,
however, is that the numerical framework needs to be entirely differentiable
because the errors in the trajectory need to be propagated through the
numerical solver itself; this idea being closely related to adjoint models in data
assimilation. Differentiable numerical models, for example in weather
and climate systems, cannot be created by simple modification of existing
climate and weather models and need to be rewritten from scratch, a complex
undertaking. Furthermore, for complex nonlinear dynamical systems, the
trajectories need to be rolled out for multiple time steps and multi-step
losses need to be optimized to ensure accuracy and stability of the learned
NN-based equations \citep{kochkov2021, keisler2022, frezat2022, list2022}; this can
substantially add to the computational burden and complexity of the training
pipeline.

Offline or \textit{a priori} learning uses high resolution numerical
models to diagnose all the SGS fluxes that are missing at lower resolutions.
Then the diagnosed SGS fluxes are directly modeled using an appropriate Neural
Network architecture which takes in the coarse-grid flow variables as inputs;
the NN-training is accomplished using standard supervised training (through
Maximum Likelihood Estimation). Methodologically this is a substantially
simpler approach and places no restrictions on the numerical solver itself as
the numerical solution and the learning of the NN are decoupled. Consequently
successfully trained NN-models can be directly incorporated into existing
climate and weather models. Correspondingly, however, a serious disadvantage
here is that the NN model has no awareness of the dynamics of the underlying system
and small errors made by the NN SGS model can accumulate when incorporated in
the numerical solver, resulting in numerical instabilities or biased solutions,
the issue becoming more acute the longer the numerical model is run
\citep{brenowitz2020, rasp2020, frezat2022}. Some of the numerical stability
issues can be alleviated through probabilistic modeling of the SGS fluxes
instead of a fixed NN-model \citep{perezhogin2023}.

Both offline and online-trained models must be tested for online or \textit{a
posteriori} stability and accuracy over justifiably long time horizons.  In
other words, the training or learning process can be operated offline (i.e.~direct
supervised learning) or online, using a differentiable numerical solver, but
all trained models must be deployed on the coarse-grained equations and tested
for online stability and fidelity. We re-emphasize the distinction between the
offline or online-learning process from actual deployment in the equations which is, by
definition, always online or \textit{a posteriori}.  Both offline and online
NN-based learning approaches have other challenges, in particular, that NNs are
optimized for Graphic Processing Units (GPUs) which most climate system models
are not currently designed to run on. In particular modern Deep NNs can have
millions of parameters and running them on CPUs can add a substantial overhead
to the numerical solver, possibly negating the effects of the gains in building
accurate SGS models.

\subsection{Related works} 
Decaying two dimensional turbulence is one of the most common model turbulent
flow problems for discovery and testing of data-driven closure and
parameterization \citep{maulik2019}. The physics of decaying turbulence is
well known \citep{brachet1988,mcwilliams1984,carnevale1991,carnevale1992} and
consists of a slow merger and interaction of vortices to form larger and larger
vortices until the entire flow domain consists of a single large vortex.
Recent studies have shown that even at high Reynolds numbers (Re$\sim$20000),
the SGS stresses in this problem can be accurately modeled with a modest number
of data samples \citep{guan2022stable, frezat2022, list2022}, leading to
accurate and stable online coarse-grid solutions at 16X or greater grid downscaling
factors.  

Continuously forced 2D-turbulence, however, is a substantially more complex
dynamical problem \citep{xiao2009} that has a persistent dual cascade of energy, an
inverse energy cascade from forcing scales to a larger scales and a forward
enstrophy cascade to smaller scales \citep{kraichnan1971}.  Only a handful of
recent studies over the past two years have successfuly obtained accurate
stable online closures for this and related problems.  \cite{guan2022} studied
the problem of a sinusoidally forced 2D turbulent flow solved in a
doubly-periodic channel, also referred to as Kolmogorov flow in classical fluid
mechanics, and demonstrated that offline learning could be effectively used to
model SGS-stress using deep CNNs (having 10-layers, about 800,000 parameters)
that resulted in stable online closures, provided sufficient amount of
training data was generated, by using around 2 million high resolution
model time-steps.  In practice they find that it is sufficient to train on only
2000 snapshots, chosen every thousand time-steps to ensure that the snapshots
were uncorrelated with each other. \cite{guan2022} also found that exploiting
the geometrical symmetry of the doubly-periodic computational domain using rotationally
equivariant convolutions substantially reduced their data requirements by a
factor of 40.  In a more recent study, \cite{ross2023} considered the offline
closure of a related idealized dynamical problem relevant to geophysical flows,
namely two-layer quasi-geostrophic flow on the $\beta$-plane; for a brief
description of the $\beta$-plane approximation, see
Sec.~\ref{subsec:eqs}. In their wide ranging study,
\cite{ross2023} examined how various choices for the inputs whether
coarse-grained velocities or velocity gradients, the form of the output
SGS-fields and the precise choice of the downscaling affected the online
accuracy of their offline-trained CNNs having 8-layers, around 300,000 parameters.

Unlike the standard approach of parameterizing the SGS stresses using NNs,
\cite{kochkov2021} parameterized the nonlinear advection term directly through
a ``neural discretization'' method demonstrated in simpler dynamical problems
in earlier papers. The \cite{kochkov2021} approach was trained in entirely
online or \textit{a posteriori} fashion through trajectory optimization. A
follow up paper \citep{dresdner2022} by the same group found accurate
online-trained parameterization of the same problem solved using spectral
methods (instead of a finite-volume approach) though using a direct
SGS-parameterizing CNN instead of the neural-discretization approach used in
\citep{kochkov2021}. A closely related study is by \cite{frezat2022} which also
employed online-learned parameterization of forced $\beta$-plane turbulence along with
standard two-dimensional turbulence, but at a substantially higher Reynolds
number than \citep{dresdner2022}. Both \cite{dresdner2022} and
\cite{frezat2022} used deep CNNs with 8-16 layers, the former work employing
the encoder-process-decoder models now gaining prominence in neural turbulent
forecasting models \citep{stachenfeld2021, keisler2022}.

\subsection{Current work}
Our approach, presented in this study, derives from the aforementioned studies
on the methodology of parameterizing sinusoidally forced two-dimensional and
geostrophic turbulence on the $\beta$-plane at high Reynolds numbers
(Re$\sim$25000) using purely offline training of the SGS stress.  However,
there are substantial departures between the choices made here which lead to
significant improvements in training and efficiency of the turbulent closures.
These points are not only technical, but also lead to new insights 
into the physical and mathematical aspects of closure problems of such
turbulent flows. We provide below a list that summarizes the main
contributions of this study.
\bi
\item[(i)] {\bf Accurate and efficient shallow CNN-closures at high
$Re$.} We demonstrate that simple \textit{two-layer} CNNs are sufficient to
obtain stable and long-term accurate online solutions at 16X downscaling for
high $Re$-turbulence. The resulting CNN-parameterizations of the SGS flux
contain only $\mathcal{O}(10^3)$ parameters;   two orders of magnitude smaller
than aforementioned recent studies,  leading in turn to substantially
faster and efficient online CNN-LES runs.
\item[(ii)] {\bf Hyperparameter space probing.} We find the best models
through extensive hyperparameter searching in the {\mk learning rate and weight
decay coefficient involved in the optimization of the loss function given by
\eqref{eq:loss} below.} This probing operation {\mk is greatly facilitated by}
the choice of small CNNs {\mk used here}. 
\item[(iii)] {\mk {\bf Optimization strategy.}} We show that the CNNs trained
using cyclical learning rate annealing result in more robust and accurate
online solutions compared to those trained using a fixed learning rate,
{\mk as traditionally operated}.
\item[(iv)] {\mk \bf Physical variables.} Our inputs to the CNN use either the
coarse velocity field, $(\bar u, \bar v)$ or the vorticity and strain fields,
$(\bar \omega, \bar \sigma_s, \bar\sigma_n)$ and outputs
are the two components of the deviatoric stress tensor; we do not consider
$(\bar \psi, \bar \omega)$ inputs because the streamfunction $\bar\psi$ is
difficult to obtain in numerical models over complex
spatial domains.
\item[(v)]   {\bf Loss function}.  Our CNNs are optimized by
minimizing  a mean square error loss function that measures their
parameterization defect with respect to the SGS expressed in terms of the
deviatoric stress tensor, and that is penalized by  the sum of squares of NNs'
weights; see  \eqref{eq:loss} below.  We do not need to use more elaborate
physics-informed loss functions that have been used in recent work
\citep{list2022, guan2022}.
\item[(vi)] {\bf Weakly non-local parameterizations and physical
implications.} The success of shallow CNNs in accurately parameterizing
turbulence for the class {\mk of turbulence problems considered here} has
direct implications for our understanding of the physics of the problem
itself. First, since the spatial non-nonlocality of CNNs grows with depth, {\mk
our closure results with shallow CNNs imply} that the SGS stresses have only a
weak non-local dependence on coarse fields. {\mk Such a weak} spatial {\mk
non-locality} of our shallow CNN-parameterizations also means that they
{\mk are more amenable for embedding} into climate models typically
relying on spatial domain decomposition techniques for computation on large
clusters; correspondingly deep CNNs would require substantially larger data
exchange between {\mk computational nodes}.

{\mk Finally, recalling that} deep NNs with a great amount of parameters are
typically required to approximate complicated nonlinear functions, due to the shallowness of our CNNs, we infer that the SGS stress must have
here a relatively simple nonlinear dependence on the coarse flow fields.  
This simple observation provides a favorable ground for the nonlinear mapping
underlying our shallow CNN-parameterization ($\Pi_{CNN}$ in
Eq.~\eqref{Eq_fundamental_relation} below) to be interpretable, which we leave
for a future work. The relative simplicity of the recent analytical operator
forms obtained by \cite{ross2023} using their hybrid genetic programming {\mk
combined with a} sparse linear regression approach is consistent with our
findings.  
\ei  

%%%%%%%%%%%%%%%%%%%%%%%%%
\section{Turbulence models and data generation}\label{sec:datagen}
\subsection{Dynamical equations and turbulent regimes}\label{subsec:eqs} 
We use a popular model fluid flow problem that exhibits complex turbulence
phenomena and is closely related to large scale turbulence in the Earth system
and in planetary atmospheres - namely sinusoidally forced two-dimensional
turbulence on the $\beta$-plane. The flow domain is specified in
$(x,y)$-coordinates and consists of a square domain of size $(L, L)$, with $L=2\pi$.
The governing equations of motion are the Navier-Stokes {\mk (N-S)} equations
which describe the evolution of flow velocity vector field, $\bv (x,y,t) =(
u(x,y,t), v(x,y,t))$, at every point in the domain {\mk and} in time.  For the
specific problem chosen here, the boundary conditions are chosen to be
periodic, {\mk namely}
\begin{align}
	u(x,y,t) =& u(x+L, y+L, t),\\
	v(x,y,t) =& v(x+L, y+L, t).
\end{align}
Note that this domain is topologically equivalent to a {\mk two-dimensional} torus.  
In two dimensions the N-S equations in the presence of background rotation, are
the evolution equations of the two components of the velocity fields,
$\bv\equiv(u,v)$ where $u$ and $v$ are the velocities along $x$- and
$y$-directions, respectively. In non-dimensional form, this becomes
\beq
	\partial_t \bv + \bv\cdot\nabla\bv +  f(\bk\times\bv) = -\mu \bv + \frac{1}{Re} \nabla^2\bv + \mathcal{\boldsymbol{F}}_{\bv}(x,y)\,.
\label{eq:NS}
\eeq
Here, $f$, the Coriolis parameter is the local rotation rate of the Earth or
some other planetary atmosphere, $\bk$ is the unit vector normal to the
$(x,\,y)$-plane while $\mathcal{\boldsymbol{F}}_{\bv}(x,y)=[-\sin(k_f y),
\sin(k_f x)]$ is the sinusoidal time-invariant forcing field that continuously
drives the flow; this specific form is chosen to be the same as in \citep{guan2022} with the
forcing wavenumber, $k_f=4$. 
The coefficient $\mu$ represents the linear drag coefficient that in
geophysical flows purports to represent the effect of bottom friction and $Re$
is the Reynolds number measuring the strength of the non-linear advection term relative to
the viscous term (the second term on the left hand
side relative to the second term on the right hand side); we choose $Re=25000$.

The velocity
field is also required to satisfy mass conservation captured through the
continuity equation, $\nabla \cdot\bv = \partial_x u + \partial_y v = 0$. The
continuity equation can be implicitly satisfied in two dimensions by defining a
scalar field called streamfunction, $\psi(x,y,t)$
\beq
u = -\psi_y\,\,,\qquad v = \psi_x 
\eeq
In two dimensions, these two equations along with the continuity constraint can
be expressed as an evolution equation of a single scalar field, the vorticity,
that is defined as the two-dimensional curl of the velocity field
\beq
\omega = \partial_x v - \partial_y u = \nabla^2\psi,
\eeq
and captures the local rotation of a fluid parcel.  The vorticity evolution
equation in two dimensions can then be written in the standard
vorticity-streamfunction ($\omega,\psi)$ form as
%%%%%%%%%%%%%%%%%%%
\beq
\partial_t \omega + J(\psi, \omega) - \beta v = -\mu \omega + \frac{1}{Re} \nabla^2\omega + F_{\omega}(x,y),
\label{eq:vort}
\eeq
where the Jacobian, $J(\psi, \omega) = \psi_x\omega_y - \omega_x\psi_y$
captures the non-linear advection term, while the vortical form of the forcing becomes 
\beq
F_{\omega}(x,y) = k_f[\cos(k_f x) + \cos(k_f y)]
\eeq
The $\beta v$-term is an approximation relevant to geophysical systems and captures the
effect of differential rotation experienced by the ocean and atmosphere in the
Earth system in a tangent plane approximation ($\beta = \frac{\partial
f}{\partial y}$). The presence of the $\beta$-term allows the flow to manifest
specific kind of waves common in planetary systems, called Rossby waves which
can substantially alter the turbulent flow dynamics relative to the flow which
results when $\beta=0$. 

Eq.~\eqref{eq:vort} is solved in Fourier space using a standard pseudospectral
method \citep{oris74}; a semi-implicit Crank-Nicholson, 2nd-order
Adams-Bashforth (CN-AB2) method is used for time-stepping with a time step of
$\Delta t = 5\times 10^{-5}$. Our baseline high-fidelity model is solved
on a square $N_0\times N_0$ grid with $N_0 = 1024$ so that our grid spacing in
physical space is $\Delta_{DNS} = 2\pi/N_0$.  According to standard parlance we
refer to these solutions as Direct Numerical Simulations (DNS) because no
parameterization is used to represent turbulence below the grid scale other
than quadratic viscosity.  Our value of viscosity coefficient, {\mk
i.e.~$1/Re$ in our non-dimensional formulation}, is chosen {\mk such that
either} decreasing the grid size by holding $Re$ fixed or increasing $Re$
(decreasing viscosity) with the grid size fixed will lead invariably to a
rapid pile-up of energy at the smallest wave numbers and eventually to numerical
instability. As a comparison, the choice of $Re\sim1000$ in
\citep{dresdner2022} allows the authors to obtain stable solutions
without any parameterization for coarser grids of up to $128\times 128$; their
problem also involves a 16X downscaling from $1024\times 1024$ to
$64\times 64$.  \cite{frezat2022} have a larger value of $Re$ than the one {\mk
used in this study} in their 32X downscaling online closure while
\cite{ross2023}, in their study in two-layer QG turbulence (for a downscaling
factor of 4X from $256\times 256$ to $64\times 64$) likely have a $Re$ value in
similar range as \cite{dresdner2022}. 
\begin{figure}[h]
\centering
\includegraphics[width=\textwidth]{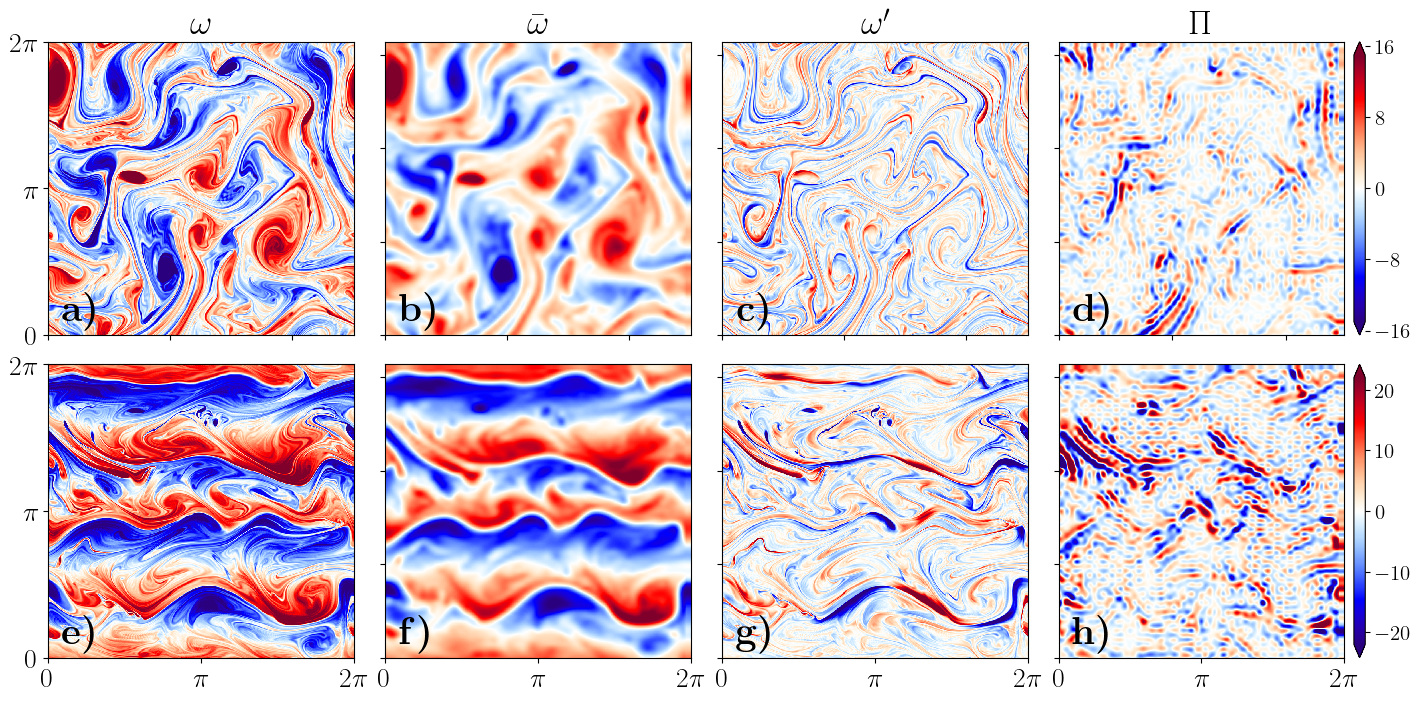} %
\hfill
	\caption{Snapshots after $t$ = 2$\times 10^6\Delta t$ for $\beta=0$ of a) DNS vorticity $\omega$
	b) Filtered DNS vorticity, $\bar \omega$ c) SGS vorticity field, $\omega' = \omega - \bar\omega$
	d) SGS stress divergence, $\Pi$.
	Same for e)-f) but for $\beta=20$; note the jets along the $x$-axis.}
\label{fig:intro}
\end{figure}
%%%%%%%%%%%%%%%%%%%%%%

In the absence of the $\beta$-term, the flow is statistically homogenous and
isotropic in space and time. Figure \ref{fig:intro}a) shows a single snapshot of
the vorticity field after $t=2\times10^6\Delta t$; large coherent vortices, typically of length scales at or larger than the forcing scale, are prominent
in a sea of fine filamentary structures. The flow kinetic energy ($\E = u ^2 +
v^2$) is concentrated at the coherent vortices as a consequence of the cascade of
energy to larger scales, while both the vortex cores and the filamentary
structures are significant reservoirs of the flow enstrophy ($\Z = \omega^2$),
the latter a consequence of the direct cascade of enstrophy to small scales
\citep{kraichnan1971}.

For finite values of $\beta$, however, the turbulent flow undergoes a
symmetry breaking instability in the $y$-direction resulting in the formation
of alternating banded zonal jets, i.e.~along $x$-direction. The mechanism of
jet formation is not a simple one and is best described as a form of turbulent
instability, called zonostrophic instability \citep{farrell2007, marston2008,
sryo2012, bakas2015}.  These jet-like structures are common in planetary atmospheres, the
most striking example being the visible banded jet structures on Jupiter.
Fig.~\ref{fig:intro}(e) shows a snapshot of the jets formed for this specific
choice of $\beta=20$ for which the coherent vortices observed in the $\beta=0$ case are no
longer visible separately but now closely interact with the jets themselves,
though the fine-scale filaments are clearly seen.
\subsection{Downscaling} 
In this section, we describe our approach to downscale or coarse-grain the
high-fidelity DNS solutions described in the previous section to a lower resolution
grid. 
Downscaling is effected in spectral space using 
a \textit{cutoff} filter, also known as a sharp spectral filter which simply 
sets modes higher than a cutoff wavenumber to be zero. For a downscaling factor of $n_d$,
the grid spacing of the reduced order model is thus $\Delta_c = n_d\Delta_0$
and the number of grid points decreases to $N_c = N_0/n_d$.  We choose $n_d =
16$, thus projecting the DNS solutions from a $1024\times1024$ grid to a $64\times 64$ grid.
In practice, before applying the cutoff filter, following \cite{guan2022}, we
first apply a Gaussian filter to the fields somewhat. The reason for this
choice is three-fold: First, \cite{zhou2019} demonstrated that this procedure
produced SGS fluxes which have a higher correlation with the coarse-grained
field, making the learning problem easier. Second, this brings the spectral
learning problem closer to finite-difference and finite-volume approach common
in Earth system models, because localized finite difference stencils can be
represented as an effective exponential cutoff filter \citep{lele92}. Finally,
only using the cutoff filter creates small scale, spatially nonlocal features
in physical space that have little physical structure and are difficult to
learn due to the spectral bias of neural networks towards learning lower
frequency more easily than higher frequency \citep{rahaman2019spectral},
especially in offline-learned parameterizations where there is no inherent
dynamical awareness; see discussion in Sec.~\ref{sec:intro}. These reasons
{\mk explain} why {\mk using} Gaussian and exponential filters {\mk became} a common {\mk practice}
before cutoff in recent studies involving CNN-based parameterizations of
two-dimensional turbulence; {\mk see e.g.~}\citep{guan2022stable, guan2022, dresdner2022,
frezat2022}.  {\mk Denoting by $\widehat\omega(k_x, k_y)$} the Fourier {\mk transform} of the vorticity,
$\omega(x,y)$, the coarse-grained field (represented
by an overbar) is written as 
\beq
\overline{\widehat{\omega}}(k_x, k_y) = \widehat{\omega}\ast G_{\Delta_c}, \;\; |k_x|, |k_y|<k_c,
\label{eq:filt}
\eeq
where $\ast$ denotes the convolution operator,  $k_c = \pi/\Delta_c = N_c/2$ is the cutoff wavenumber (for $N_c=64$, $k_c = 32$) and the {\mk Gaussian filter $G_{\Delta_c}$ is given by 
\begin{align}
G_{\Delta_c}(k_x, k_y) &= \exp\bigg(-\Big(\frac{k_x^2 + k_y^2}{24}\Big)\Delta_F^2\bigg)\label{eq:filtgauss}\\
		       & = \exp \bigg(-\frac{\pi^2}{6}\frac{k^2}{k_c^2}\bigg),
\end{align}
 with $\Delta_F = 2\Delta_c$; see \citep{guan2022stable}.} The
fields filtered from the DNS solutions {\mk on} the lower-resolution grid
are referred to as the Filtered DNS (FDNS) fields. Fig.~\ref{fig:intro}b) and
f) display snapshots of the 16X downscaled fields (i.e.~the FDNS fields)
corresponding to the DNS snapshots in Fig.~\ref{fig:intro}a) and e) for the
cases of $\beta=0$ and $\beta=20$ respectively. Interestingly, \cite{ross2023} find that
the above Gaussian filtering approach makes the learning problem more challenging but we find few issues
with either the offline learning of the CNN or its online deployment.
%%%%%
\subsection{Sub-grid scale stresses}
We apply the filtering step in Eq.~\eqref{eq:filt} to the momentum equation
\eqref{eq:NS} and the corresponding vorticity equation \eqref{eq:vort}. The
momentum equation is then rewritten as
\beq
 \partial_t \bar\bv + \bar\bv\cdot\nabla \bar\bv +  f(\bk\times\bar\bv) = -\mu \bar\bv + \frac{1}{Re} \nabla^2\bar\bv + \mathcal{\boldsymbol{F}}_{\bv}(x,y) + \nabla\cdot\bar\bS,\\
\label{eq:filtNS}
\eeq
where the SGS momentum flux tensor (sometimes just referred to as the stress tensor) is
\beq
\bS=   \begin{bmatrix}
	\tau_{uu}  &  \tau_{uv}\\
    \tau_{uv}         & \tau_{vv}
    \end{bmatrix}\,,
    \label{eq:stress0}
\eeq
where $\tau_{uu} = \overline{u^2} - (\bar u)^2$, $\tau_{uv} = \overline{uv} -
\bar u\bar v$ and $\tau_{vv} = \overline{v^2} - (\bar v)^2$ are the three
tensor components as $\bar\bS$ is clearly a symmetric tensor. 
{\mk We decompose Eq.~\eqref{eq:stress0} as} 
%%%%%%%%%%%%%%%%
\beq
\bS=   \underbrace{\begin{bmatrix}
	\frac{\tau_{uu} - \tau_{vv}}{2}  &  \tau_{uv}\\
	\tau_{uv}         & -\frac{\tau_{uu} - \tau_{vv}}{2}
\end{bmatrix}}_{\bS_d}
	\,+\,	
	\underbrace{\frac{\tau_{uu} + \tau_{vv}}{2}\bI}_{\bS_0}\,,
    \label{eq:stress}
\eeq
where $\bI$ is the identity matrix and $\E = \frac{\tau_{uu} + \tau_{vv}}{2}$
is the SGS kinetic energy. The first term on the {\mk right-hand side (RHS)}, $\bS_d$ is referred to
the deviatoric stress tensor and has only two independent components with the
third independent component now appearing as the magnitude of the diagonal
tensor, $\bS_0$. Note the forcing terms are not affected by the filtering
operation as we always chose our cutoff scale, $k_c>k_f$.

Applying the filter, Eq.~\eqref{eq:filt} {\mk to} the vorticity equation,
Eq.~\eqref{eq:vort}, we arrive at the ``filtered'' equation corresponding to
\eqref{eq:filtNS} in the form
\beq
\partial_t \bar\omega + J(\bar\psi, \bar\omega) - \beta \bar v = -\mu \bar\omega + \frac{1}{Re}  \nabla^2\bar\omega + f(x,y) + \Pi, 
\label{eq:vortFilt}
\eeq
where 
\beq
\Pi = \overline{J(\psi, \omega)} - J(\bar\psi, \bar\omega),
\label{eq:form1}
\eeq
is the SGS vorticty flux \textit{divergence} that can be written in two other
alternative forms. The first form expresses $\Pi$ in terms of the divergence of
the SGS vorticity flux, \beq
\Pi=\nabla\cdot(\overline{\bv\omega}-\bar\bv\bar\omega).
\label{eq:form2}
\eeq
The second form, more directly relevant to the approach {\mk pursued} in this manuscript, relates $\Pi$ to the the SGS momentum flux tensor, 
$\bS$ 
\beq
\Pi = \nabla\times(\nabla\cdot\bS) = \nabla\times(\nabla\cdot\bS_d)
\label{eq:form3}
\eeq
where the curl operation above is the two dimensional curl. The second equality
above results because $\bS_0$ is a diagonal tensor, and it follows that
$\nabla\times(\nabla\,\cdot\,\bS_0)=0$. Fig.~\ref{fig:intro}d) and h) show snapshots
of $\Pi$ for the cases $\beta=0$ and $\beta=20$ respectively; the corresponding SGS vorticity
fields, $\omega' = \omega - \bar\omega$ are also shown in the same figure for reference. There is
a great deal of correlation between $\omega'$ and $\Pi$ but this is, evidently, not a simple dependence.

Having a data-driven model of $\Pi$ allows us, in principle,  to solve Eq.~\eqref{eq:vortFilt}
though the question arises as to which form of $\Pi$ above should be used for
modeling. The most common form used is the one in {\mk Eq.~\eqref{eq:form1}
where} $\Pi$ is directly modeled using a neural network
\citep{maulik2019,guan2022, guan2022stable, frezat2022}. An alternative
approach might be to model the two components of the SGS vorticity flux vector
$\overline{\bv\omega}-\bar\bv\bar\omega$ and to compute the divergence to
compute $\Pi$ in Eq.~\eqref{eq:form2}. The approach adopted here {\mk is
inspired by} \citep{zanna2020}, {\mk namely to} model the components of the SGS
momentum flux, $\bS$ and compute $\Pi$ using Eq.~\eqref{eq:form3}.
\cite{zanna2020} used a single convolutional neural network to model the three
components of $\bS$, namely $\tau_{uu}$, $\tau_{vv}$ and $\tau_{uv}$.  We propose in this study a variation of this approach by learning a CNN
approximation of the deviatoric stress tensor's two components, $(\tau_{uu} -
\tau_{vv})/2$ and $\tau_{uv}$,  which, as pointed above, is sufficient to obtain
$\Pi$; see Eq.~\eqref{eq:stress}.  

%%%%%
\subsection{The locality of the resolved and SGS scale interactions}
\begin{figure}[h]
\centering
\includegraphics[width=\textwidth]{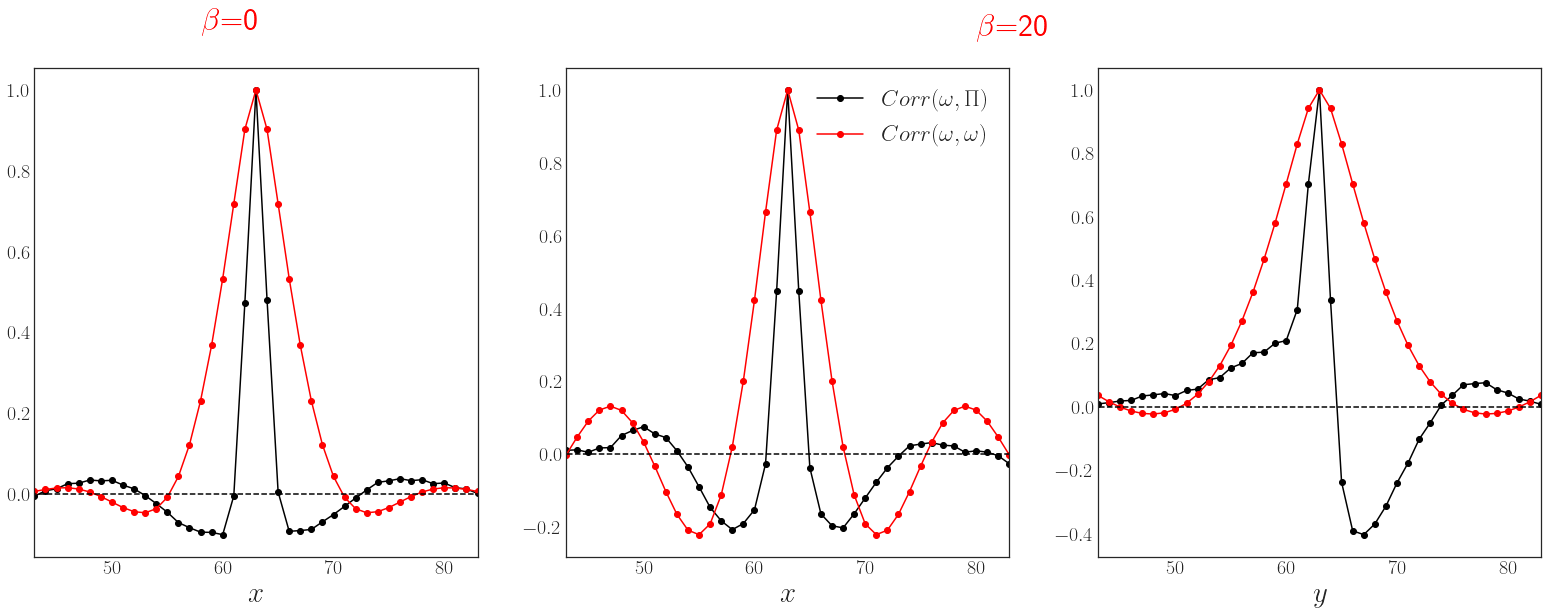} %
\hfill
	\caption{$x$- and $y$-sections of the spatial cross-correlations between $\Pi$ and the resolved vorticity field, $\bar \omega$. For $\beta=0$, the correlations along both directions are nearly symmetric so only one of them is shown. For finite $\beta$,
the cross-jet correlation length (along $y$) is longer than the along-jet correlation length that is comparable to that in the $\beta=0$ case.}
\label{fig:corr}
\end{figure}

The interactions between the resolved fields and the SGS motions are only
weakly non-local \citep{eyink2005, eyink2009}. This property can be e.g.~inferred by
examining the two-dimensional spatial cross-correlations between 
$\bar\omega$ and $\Pi$. For reasons of efficiency the spatial cross-correlation
is computed for each snapshot by using fast Fourier transforms through
the cross-correlation theorem (a generalization of the Wiener-Khinchin theorem)
\citep{fisher2008}, and then averaged across time.  In Fig.~\ref{fig:corr}, we
highlight the $x$- and $y$-sections of the cross-correlation. When
$\beta=0$, the cross-correlation is essentially isotropic and examining any of
these sections suffices; it can be observed in Fig.~\ref{fig:corr} that the
correlation is weak beyond a 7-point region in space. When $\beta=20$ the
along-jet correlation mirrors the $\beta=0$ result but the cross-jet
correlation length is larger; this relates to interactions and coupling between
the jets composing the flows. Even in this case, however, highly correlated regions are
still relatively local. This observation motivates the choice of our neural network
architecture employed in this manuscript, as detailed subsequently.

%%%%%%%%%%%
As a side remark, we emphasize that autocorrelation of the resolved vorticity
highlights the average size of the eddies in the $\beta=0$ case and the jet size when $\beta=20$ (red curves
in Fig.~\ref{fig:corr}). 
%%%%%%%%%%%%%%%%%%%%%%%%%%%%%%

%%%%%%%%%%%%%%%%%%
\subsection{Galilean invariant SGS models} \label{subsec:galie}
The equations of motion, Eqns.~\eqref{eq:NS} and \eqref{eq:vort} are Galilean invariant, i.e. 
invariant to changes in the inertial frame of reference. To demonstrate this, and without loss of generality, we can 
re-write the equations in a reference frame translating with constant velocity
(denoted by primes), $\bV\equiv(U, V)$ along $x$ and $y$ directions, i.e.
\beq
x' = x -Ut, \quad y' = y - Vt, \quad t' = t. 
\label{eq:inertial}
\eeq
and
\beq
u' = u + U,\quad v' = v + V,\quad \psi' = \psi + Vx - Uy,\quad \omega'=\omega.
\label{eq:galilean}
\eeq
Note that the vorticity is invariant to a uniform translation while the other quantities are not; this is because the vorticity
is comprised of gradients of velocity; in general all four possible gradients of velocity are invariant to Galilean transformations
\beq
v_x' = v_x,\quad v_y' = v_y,\quad u_x' = u_x,\quad u_y' = u_y,
\eeq
as are linear combinations of these velocity gradients, including $\omega$ and the two strain components, the normal and shear strains defined respectively as,
\beq
\sigma_n = v_x-u_y, 
\eeq
and
\beq
\sigma_s=v_x+u_y.
\eeq
%%%%%%%%%%%%%%%%%%
An important consequence of Galilean invariance is that the filtered equations \eqref{eq:filtNS} and \eqref{eq:vortFilt} must
also be Galiean invariant (the filtering operator being independent of time) and consequently so must be the SGS fluxes, $\bS$ and $\Pi$.

Turbulence modeling or closure refers to the modeling of SGS fluxes, either
$\Pi$ directly or modeling $\bS$ and then using Eq.~\eqref{eq:form3} to obtain
$\Pi$, as functions of the coarse field variables, although different choices for the
input variables are possible  as mentioned above. A natural choice in solving the filtered vorticity
equation \eqref{eq:vortFilt} is to choose the input coarse field variables to be 
 $(\bar\psi,\bar\omega)$. This choice made in \citep{maulik2019, guan2022stable, guan2022, frezat2022} consists of $\Pi$ modeled directly as
$\Pi=\Pi_{\theta}(\bar\psi, \bar\omega)$, in which $\theta$, based on common parlance, represents the set of parameters
comprising the model which can range from a simple linear regression model to
more complex choices like neural networks. Alternatively, one may choose the coarse-grained velocities
as inputs \citep{dresdner2022}, i.e.
$\Pi=\Pi_{\theta}(\bar u, \bar v)$.

\cite{zanna2020} use $(\bar u, \bar v)$ as inputs {\mk and} model the three components of $\bS=\bS_{\theta}(\bar u, \bar v)$ to
obtain $\Pi$ through Eq.~\eqref{eq:form3}. 
However, it should be noted that none of these choices are Galilean invariant, because on changing the 
intertial frame of reference leads us to changes in $(\bar\psi, \bar u,\bar v)$ [as per Eq. \eqref{eq:galilean}]
\beq
\bar u' = \bar u + U,\quad \bar v' = \bar v + V, \quad\bar\psi' = \bar\psi + Ux + Vy.
\eeq
thus \textit{a priori}, models that are functions of the above variables will not be Galilean invariant 
though given sufficient data, the NN with sufficiently high
number of parameters would likely learn this  
physical
symmetry. However, it is possible to make a simple but sufficient modeling choice that implicitly assumes Galilean invariance.
For the two components of deviatoric stress tensor, $\bS_d$, {\mk we aim at finding the following parameterization}
\beq
{\mk \bS_d=\bS_d^{\theta}}(\bar\omega, \bar\sigma_n,\bar \sigma_s ),
\label{sgs:vortstrain}
\eeq
where $\theta$ represent the set of all parameters of the CNN.
Because $(\bar\omega, \bar\sigma_n,\bar \sigma_s )$ represent all possible linear combinations of gradients of the velocity,
adding a uniform velocity leaves them trivially invariant\footnote{Note that the fourth linear combination, the divergence, $\bar \delta=\bar u_x + \bar v_y= 0$ due to the continuity relation.}.
We also learn {\mk below} non-Galilean invariant models, following \cite{dresdner2022, zanna2020, ross2023} as
\beq
\bS_d=\bS_d^{\theta}(\bar u, \bar v ).
\label{sgs:uv}
\eeq
We choose not to learn {\mk neural} models that use as inputs
$(\bar\psi,\bar\omega)$ (as e.g.~in \cite{guan2022stable, frezat2022}) for the
simple reason that the streamfunction is generally not available in realistic
geophysical models, making such choices not easy to generalize {\mk in
practice}. {\mk It is indeed important to note that on} the spectral plane {\mk
the differential operators involved are often} diagonal, but in general a
solution of an expensive Poisson equation is required to obtain $\psi$.
\subsection{Baseline parameterizations}
Following recent studies \citep{guan2022, frezat2022, ross2023}, we choose
the parameterizations by Smagorinsky and Leith as baselines. These models see
widespread use in both contemporary LES studies and realistic numerical models
currently being used in the climate system \citep{bachman2017Leith,
pearson2017}.  The Smagorinsky parameterization is generally implemented
in the momentum equations, \eqref{eq:filtNS}, and parameterizes $\bS_d$ in terms of the coarse
strain tensor. However, we choose the version of Smagorinsky that is directly
implemented in terms of the coarse vorticity field, $\bar\omega$
\citep{maulik2019} and in divergence form \citep{frezat2022}; note that the
momentum and vorticity forms of Smagorinsky are not equivalent. Both the above
parameterizations are \textit{diffusive} parameterizations and assume that
$\Pi$ can be represented in diffusive form as
\beq
\Pi = \nabla\cdot(\nu_e\nabla\bar\omega),
\eeq
where the only unknown is the `eddy' diffusivity, $\nu_e$.
The Smagorinsky parameterization models $\nu_e$ as 
\beq
\nu_e =  c_s\Delta_c^2|\bar S|,
\eeq
where $|\bar S|^2 = \sigma_n^2 + \sigma_S^2$ is the strain magnitude and $\Delta_c$ is the grid size of the coarse grid. The Leith paramaterization takes the form
\beq
\nu_e =  c_l\Delta_c^3|\nabla\bar\omega|\,. 
\eeq
Note that both of the above parameterizations are Galilean invariant as explained
in the preceding section.  Following \citep{ross2023}, we vary the constants
$c_l$ and $c_s$ in $[0.01,1.0]$ and run coarse-grid solutions with Smagorinsky and
Leith parameterizations to similarly long times as the CNN-LES to allow for a
direct comparison. As a foreshadowing of the results ahead, we note that in the vorticity
form expressed above, the flows obtained from the Leith and Smagorinsky models are extremely similar
when $c_s=c_l$.
%%%%%%%%%%%%%%%%%%%%%
\section{Machine learning framework}
\subsection{CNN architecture and non-locality}\label{subsec:arch}
\begin{figure}[h!]
\centering
\includegraphics[width=.9\textwidth, height=.4\textwidth]{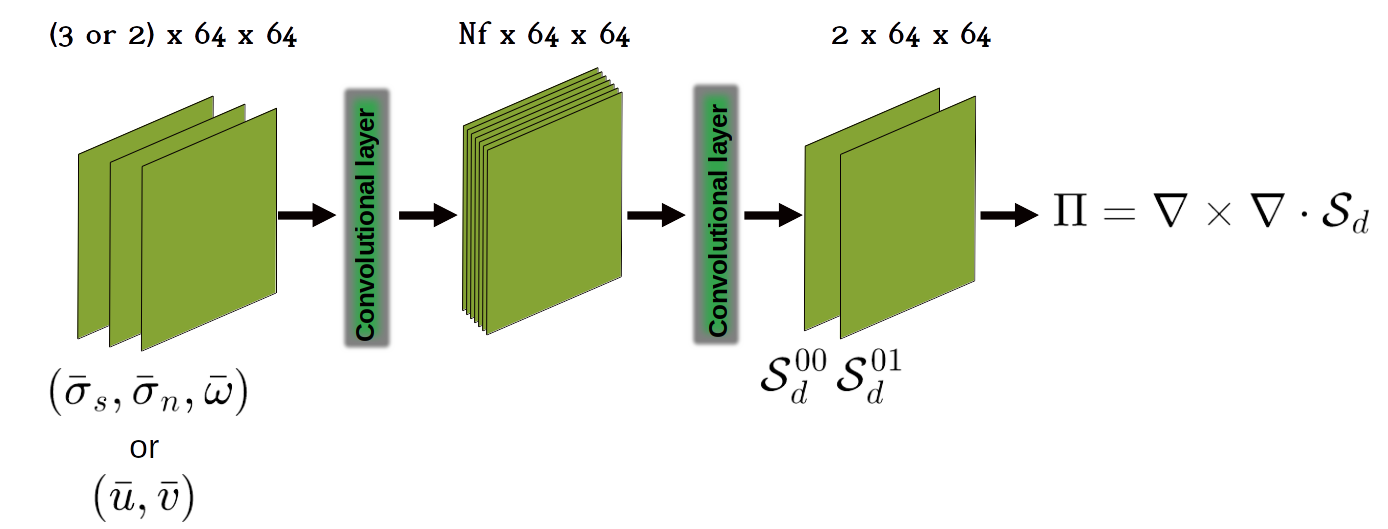} %
\hfill
  \caption{{\bf Pipeline of CNN used to model SGS stresses.} The input fields
are either of the coarse field combinations $(\bar\omega, \bar\sigma_n,
\sigma_s)$ or $(\bar u, \bar v)$ while the output filters are the two relevant
components of the deviatoric stress tensor, $\bS^{00}_d$ and
$\bS^{01}_d$. The SGS vorticity fluxes are computed first
through a tensor divergence followed by a two-dimensional curl operation. 
}
\label{fig:2layercnn}
\end{figure}

We employ an entirely offline machine learning pipeline to learn the SGS fluxes
in this manuscript. As described previously, to solve the filtered vorticity
equation, Eq.~\eqref{eq:vortFilt}, we model {\mk $\bS_d$} using a Convolutional
Neural Network that takes in the filtered field variables as inputs, either
$(\bar u, \bar v )$ or $(\bar\omega, \bar\sigma_n,\bar \sigma_s )$; $\Pi$ is
then obtained from {\mk $\bS_d$} using Eq.~\eqref{eq:form3}.  Convolutional layers are
translationally-equivariant and a natural choice because the Navier-Stokes
equations also have the same symmetry. {\mk Recent} studies
\citep{guan2022stable, guan2022, dresdner2022,frezat2022,ross2023} employ deep CNNs
for parameterization consisting of a few hundred thousand to a million
parameters. We find, however, that \textit{shallow} CNNs, in particular, simple
two-layer neural networks with only {\mk $\mathcal{O}(1000)$} parameters are
sufficient to model SGS stresses. Our precise architecture is shown in
Fig.~\ref{fig:2layercnn}.  The input layer consists of 2 or 3 inputs depending
on whether the velocities or combination of vorticity and strains are chosen as inputs; the
sole hidden layer consists of $N_f$ convolutional filters with the two output layers
only having the two independent fields that constitute {\mk $\bS_d$}, namely 
$({\mk \bS_d^{00}, \bS_d^{01}})=(\frac{\tau_{uu} - \tau_{vv}}{2}, \tau_{uv})$;
{\mk see Eq.~\eqref{eq:stress}}.  The choice of the filter width, commonly
referred to as the kernel size, of each convolutional filter in each layer is
set to $ks = 5$. $N_f$ is a key hyperparameter, {\mk which we choose to be
either $8$, $16$ or $32$}. The total number of parameters for each of
these choices is 800, 1600, and 3200 for the case of $(\bar u, \bar v)$ input
respectively. In either case, the number of convolutional filters is
unusually low compared to the norm; our objective being to find the
smallest possible NNs which {\mk lead to accurate and stable} solutions and we
show that these actually suffice. We employ a single {\it swish activation
function}, $\mathcal{S}$,  after the first layer; i.e.~$\mathcal{S}(\xi) =
\xi H(\xi)$ where $H(\xi)$ is the sigmoid function. Swish tends to work better
than ReLU across a number of challenging data sets
\citep{ramachandran2017searching}.

Two important points underline our choice of shallow CNN, the first of which is
the \textit{effective non-locality}. This quantity is also known as the
receptive field (RF) of a CNN in the computer vision literature.  It is defined as the
set of points in the input domain which is connected to the center of the point
region that the CNN is currently operating on. Alternatively this is the
effective filter width of the CNN. For example, a single convolutional layer
with a 5$\times$5 filter has a trivial 5$\times$5 = 25 point RF, {\mk i.e.~an
effective filter width of 5}. Because NNs are compositions of layers comprising
convolutional filters, the RF is a linear function of the number of filters.
Thus a two-layer CNN with 5$\times$5 filters has {\mk an} RF of 9$\times$9
points, i.e.~an effective filter width\footnote{Note that for a
general $n$-layer CNN comprised of $ks=5$ filters, its effective width is $2n-1$.} of 9. 
{\mk As a comparison, the} 10-layer CNN employed by
\cite{guan2022} has an RF of 41$\times$41, which on a 64$\times$ 64 domain is
nearly global in its non-locality. The study by \cite{maulik2019} used
spatially local NNs, in the form of a fully connected NN instead
of CNN, with a receptive field only equal to $3 \times 3$; too low for the current problem.
\cite{ross2023} employed an 8-layer CNN with a RF of 21$\times$21; they
examined the gradient of the output SGS fluxes with respect to the inputs at a
single point for their $\beta=0$ case and found non-zero values in only
a 9$\times$9 spatial region, consistent with the RF of our 2-layer CNN chosen in
this study. However, we also find the non-trivial result that the same 2-layer
CNN architecture also suffices in accurately modeling the $\beta=20$ jets case
as demonstrated in subsequent sections.

 The RF is closely connected to the \textit{stencil size} in numerical
discretization and can be directly identified as such in the context of solving
PDEs.  {\mk While our choice of CNNs assumes implicitly non-locality as in
other studies of SGS closure, in contradistinction, our closure results with
2-layer CNNs presented hereafter allow us to claim that the inherent degree of
non-locality in SGS parameterization is weak, even at high Reynolds number.}
This observation has  implications for deployment of
SGS-parameterizations in ocean and atmospheric models which often use a spatial
domain decomposition for solution on large compute clusters. Having a CNN with
large spatially locality would pose severe limitations on the domain
decomposition, which are alleviated by choosing shallow CNNs.  

The second point of note is the effective non-linearity. Due to their
hierarchical structure, deeper NNs are able to represent a more complex family
of nonlinear functions than shallow NNs having the same number of parameters
\citep{bengio2011}. Thus lower layers in the network learn simple features,
while higher layers build upon these to represent more complex patterns and
abstractions. Therefore our choice of 2-layer CNN implicitly makes certain
assumptions towards simplicity and sparsity of the relationships between the
SGS fluxes and the input coarse flow variables. It is possible to increase the
degree of nonlinearity of the CNN without increasing its RF or non-locality
through 1$\times$1 ``convolution'' layers which are strictly speaking not convolutions but fully connected layers across the feature dimension. However, we do not find
a need for such choices.

%%%%%%%%%%%%%%%%%%%%
%%%%%%%%%%%%%%%%%%%%%%%%%%%%

\subsection{Data generation and loss function}
Since the output of our CNN is the deviatoric stress tensor, $\bS_d$, we use
Eq.~\eqref{eq:form3} to compute the SGS vorticity flux divergence used in
Eq.~\eqref{eq:vortFilt}. Our loss function is a simple $L_2$-loss between the
{\mk CNN-parameterization of $\Pi$} and that obtained from FDNS, {\mk that is
regularized as follows} 
%%%%%%%%%%%%%%%%%%%%%%%%%%%%%%
\beq\label{eq:loss} 
\L = {\mk \|\nabla\times (\nabla
\cdot {\bS_d^{\theta}) - \Pi_{\textrm{FDNS}}\|_2^2} + \text{wd}\,\sum
\theta_j^2}.  
\eeq 
%%%%%%%%%%%%%%%%%%%%%%%%%%%%%%
{\mk Here, the norm $\| \cdot \|_2$ refers to the $L_2$-norm over}
the flow domain {\mk while} the $L_2$-penalty {\mk coefficient} {\mk on the sum
of squares of all the weights of the NNs} is referred to as the ``weight
decay'' {\mk coefficient}. Compared to recent studies \citep{list2022,
guan2022}, the use of additional physics-informed losses to further constrain
our offline training, did not turn out to be an important ingredient to derive
accurate closures within our  shallow CNN framework. As shown in subsequent
sections, we find high fidelity solutions without recourse to adding
more physical constraints as part of the loss function.

To train our 2-layer CNNs, we generate four separate trajectories of high resolution
solutions with differing random initial conditions.  Following {\mk the protocol of}
\cite{guan2022}, we use weakly correlated snapshots that are 1000$\Delta t$ iterations
apart for training and testing, to promote diversity within the training dataset.  We use 200 snapshots each from the first two
trajectories as our training set with a total of 400 snapshots corresponding to 400,000
high-resolution model iterations, and 400 snapshots of the third trajectory for
the test set and the fourth trajectory (the validation trajectory) to initialize
our CNN-LES run and compare with the corresponding FDNS evolution. Since we
wish to characterize the fidelity of our parameterized solutions over long time
horizons, we run the fourth trajectory for $T=6\times10^6\Delta t$, which is also
the time for which we run our online CNN-LES solutions for validation.

A minor if important detail here is that we do not normalize our input or
output data before applying the CNN on it, nor do we use any normalization
layers like Batch Normalization \citep{ioffe2015}.

\subsection{Learning rate annealing and model selection}
\begin{figure}[h]
\centering
\includegraphics[width=\linewidth]{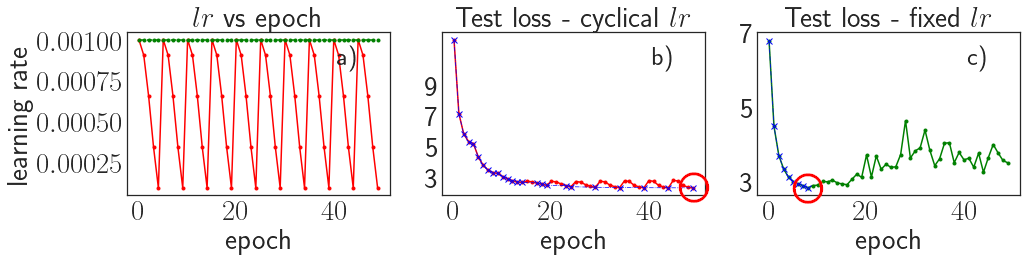} %
\hfill
	\caption{Left: The learning rate of cosine annealing case (red) vs the
	fixed case (green) as a function of epochs. Middle: The total test loss
	(red) with the lower bound curve of test loss (blue) for the cyclical
	annealing rate case.  Note that the blue curve is a monotonically
	decreasing function. Right: The same as the middle panel but for the
	fixed $lr$ case. The red circle shows the model state that has the lowest test 
	error over the entire course of training; the CNN model is saved for this point
	and used to drive the online CNN-LES runs.
}
\label{fig:lr}
\end{figure}
%%%%%
The learning rate is crucial hyperparameter in the optimization of neural
networks. While the simplest training approach is for the learning rate to be
kept constant through the entire training process, specific forms of
$lr$-annealing can substantially improve test accuracy. We choose cosine
annealing with warm restarts \citep{loshchilov2016} because of its tendency to
provide a strong implicit regularization and robust solutions across a wide
range of hyperparameters. 

Figure \ref{fig:lr}(a) shows the cyclically annealed $lr$ as a function of
epoch. {\mk Within this optimization framework, $lr$ is initiated at} a maximum value ($lr$=0.001 in
Fig.~\ref{fig:lr}) and  is decreased to zero via the cosine function over the course
of 5 epochs before suddenly ramping back to its original value. Typical test
losses as a function of epoch are shown in Fig.~\ref{fig:lr}b (for the
annealed case, red curve) and  Fig.~\ref{fig:lr}c (for the fixed $lr$ case, green curve). 
For the annealed case, after the initial decrease, the test loss also rises 
%and falls with the $lr$-cycle, 
but reaches a new  local minimum after each cycle,  {\mk with smaller loss function values than previously reached.}

For the fixed-$lr$ case, however, the test loss decreases to a minimum
value at which the CNN model is chosen as the optimal one but then becomes
erratic as the number of epochs increases.  We also show the lower-bound envelope of the test loss for each case
(in blue); note that this curve is monotonically decreasing over a larger number of epochs in the cyclically annealed case.  In each case, we select the
CNN model with the least test loss over the the entire duration of training which
corresponds to the right-most point on the blue curve as marked by the red circle
in Fig.~\ref{fig:lr}b-c.  Because of our specific model selection criteria, we
do not need to monitor our training process, a consequence of which is that our
test set needs to be diverse enough to reflect the dynamics' variability as well as as 
sufficiently large. 
%%%%%%%%%%%%%%%%%%%%%%%%%%%%%%%%%%

\subsection{Hyperparameter grid search}\label{Sec_hyperparam_search}
The {\mk other two central} hyperparameters in the model optimization process are the {\mk batch
size and weight decay}. Typical batch sizes are chosen to be
larger than {\mk some threshold}, but we choose the batch size to be \textit{unity} for all our
model training which normally would lead to extremely slow and noisy training.
The main reason here is that, as explained in Sec~\ref{subsec:arch} {\mk above}, the 2-layer
CNN chosen here has an effective non-locality of 9$\times$9 which is the region in the
coarse-grid domain where the CNN acts on independently of the other parts of
the domain.  Since our coarse-grid domain has size, 64$\times$64, this means that our
\textit{effective} batch size is actually $64^2/9^2\approx 50$.
{\mk For sufficiently deep CNNs the effective batch size and actual batch size would be identical.}

To find the  best CNN parameterizations for our closure, we perform an
extensive hyperparameter search in $(lr, wd)$ search space. For each value of
the hidden layer size, {\mk $N_f$ in $\{8, 16, 32\}$ we vary our learning rate
$lr$ in $\{10^{-4}, 10^{-3}, 10^{-2}\}$ and {\mkr our weight decay coefficient}, $wd$, in $\{10^{-5}, 10^{-4},
10^{-3}\}$.} For each of these parameter choices we train models for both fixed
$lr$ and cyclically annealed $lr$ where the chosen value of $lr$ for the
annealed case represents the maximum value of $lr$.  For each of the above
choices, we consider two classes of inputs, $(\bar u, \bar v)$ and $(\bar
\omega, \bar \sigma_n, \bar \sigma_s)$ as described in Sec.~\ref{subsec:galie}.
Thus we train and test a total of $ 3^3\times2\times2=108$ CNN models ($2$
types of optimizations---annealing or not; 2 types of inputs) each for the
$\beta=0$ and for $\beta=20$  cases. Note that because our CNNs have such a
small number of parameters, both our training times and CNN-LES runtimes are
extremely fast. 

Training each CNN model to 50 epochs takes only about 1.5 minutes on a V100 GPU
while running the CNN-LES model {\mk for} $T = 6\times10^{6} \Delta t$ takes about
20 minutes. Due to the 16X downscaling adopted here,  the time-step of
our downscaled run is 16 times larger so this is equivalent to around 375,000
$\Delta t_{LES}$. We note that 84 of the 108 cases when $\beta=0$ and 87 of the 108 cases when
$\beta=20$ are numerically stable through the course of the entire online run. 
We also observed two other solutions which became numerically unstable after
$t$=5$\times10^6\Delta t$ illustrating the challenges with evaluating online stability.

%%%%%%%%%%%%%%%%%%%%%%%%%
\subsection{Diagnostics}\label{sec:diag}

We evaluate the accuracy and fidelity of our online CNN-LES runs using a variety of
metrics which quantify the structure and dynamics of the CNN-parameterized
downscaled equations, relative to the ground truth FDNS solution. The kinetic
energy spectrum, $\hat{E}(k)$, where the $\hat{\cdot}$ symbol
represents that the quantity is in spectral space and $k^2 = k_x^2 + k_y^2$ is the
radial wavenumber, is a fundamental metric for quantifying turbulent flows and dynamical regimes
spanning spatial scales.

We can also write more dynamically relevant cross-scale kinetic energy and enstrophy
fluxes associated with the SGS flux, defined in spectral space as
\begin{align}
E_{flux}(k) &= \mathfrak{R}\left(-\Pi(k)^*\hat{\bar \psi}\right)\,,\\
Z_{flux}(k) &= \mathfrak{R}\left(-\Pi(k)^*\hat{\bar \omega}\right)\,, 
\end{align}
where $\mathfrak{R}$ denotes the real part,  ${\Pi(k)}$ the SGS vorticity flux
divergence in spectral space and {\mk $z^*$} represents the complex conjugate of $z$.\footnote{Note that to simplify notations, we have dropped the $\widehat{\cdot}$ symbol over the energy, enstrophy and $\Pi$ symbols.} 
Note that the above fluxes do not represent the total flux of kinetic energy and
enstrophy in the CNN-LES solutions but only the contributions associated with
$\Pi$.  $\Pi(k)$ is itself a metric which can be compared between the FDNS and
corresponding {\mk CNN-LES} runs as a test of the fidelity.

We also construct the probability distribution functions of the coarse
vorticity field, $P(\bar\omega)$ and the SGS flux, $P(\Pi)$. For turbulent
flows, these quantities are not Gaussian and can have strong tails, a problem
exacerbated by the intermittency of two-dimensional and geostrophic turbulence
caused due to the persistence of long lived vortices. The intermittency problem
ensures that without extremely long time simulations, the tail of the
distributions are difficult to capture accurately and remain noisy.  In general
the standard approach for computing these quantities for turbulent flows is
through Kernel Density Estimation (KDE) \citep{guan2022, ross2023}, 
which depends on the the choice of the underlying kernel and method used to obtain the best fit distribution \citep{Botev_al10}.
However, by
choosing to run our online validation solutions {\mk over} long time horizons {\mk of size} $T = 6\times 10^6 \Delta t$, we obtain well defined probability density function (PDF) estimates  by simply estimating histograms without recourse to KDE. We compute our spectra, fluxes
and PDFs for the online CNN-LES solutions and the corresponding FDNS solution
using 3000 snapshots separated by $2000\Delta t$.

{\mk For each of these spectral, fluxes and PDF quantities,}  we define metrics {\mk aimed at evaluating similarities} between the FDNS
and CNN-LES solutions. {\mk The choice of these} metrics are 
related to those used by \cite{ross2023}.  For the energy fluxes, we simply use
the coefficient of restitution between the FDNS (superscript $^D$) and online 
CNN-LES (superscript $^{\theta}$) quantities
to measure the disagreement as
\begin{align}
\texttt{\small energy-flux-diff} &= 1 - R^2\left(E_{flux}^{D}(k), E_{flux}^{\theta}(k)\right),\\
\texttt{\small enstrophy-flux-diff} &= 1 - R^2\left(Z_{flux}^{D}(k), E_{flux}^{\theta}(k)\right),
\end{align}
The energy and SGS divergence spectra of the online solutions are evaluated in
similar fashion, though we replace the quantities themselves with their
logarithm to ensure that errors at higher wavenumbers  are adequately
represented. Thus, 
%%%%%
\begin{align}
\texttt{\small spectral-diff} &= 1 - R^2\left(\log[E^{D}(k)], \log[E^{\theta}(k)]\right),\\
\texttt{\small spectral-sgs-diff} &= 1 - R^2\left(\log[\Pi^{D}(k)], \log[\Pi^{\theta}(k)]\right),
\end{align}
%%%%%%%
Metrics that compare probability distributions are often referred to as divergences and a variety
of options exist. We choose an $L_2$-divergence which is simply the integrated square difference between
the PDFs of the FDNS and online CNN-LES distributions,
\begin{align}
\texttt{\small distrib-diff} &= \int_{-\infty}^{\infty}[P(\omega^{D}) - P(\omega^{\theta})]^2\dd\,\omega,\\
\texttt{\small distrib-sgs-diff} &= \int_{-\infty}^{\infty}[P(\Pi^{D}) - P(\Pi^{\theta})]^2\dd\,\Pi.
\end{align}

%%%%%%%%%%%%%%%%%%%
\section{Results}\label{sec:results}
\subsection{Offline accuracy}\label{sec:offline}
\begin{figure}[!h]
\centering
\includegraphics[width=\linewidth]{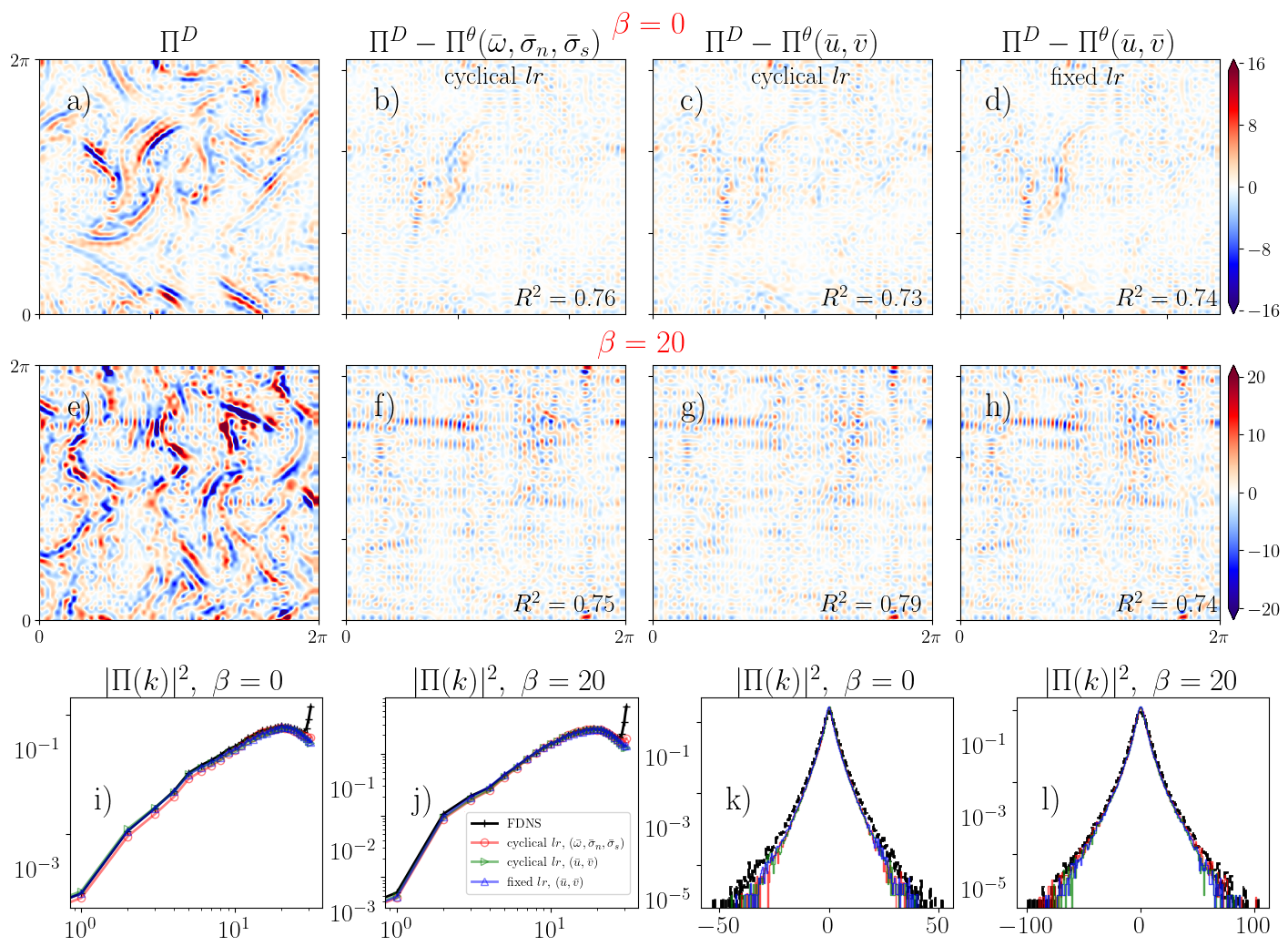} %
\hfill
	\caption{A comparison of the test data SGS flux divergence, $\Pi^D$
with that predicted by the trained CNN models, $\Pi^{\theta}$ having different
inputs and training methodologies.  {\bf Top row}: Comparison of a single
test set snapshot, $\Pi^D$ with the the error of the CNN predictions $\Pi^D -
\Pi^{\theta}$ depending on whether the inputs are either $(\bar u, \bar v)$ or
$(\bar\omega, \bar\sigma_s, \bar\sigma_n)$ or whether cyclical learning rate
annealing was used for training, for $\beta=0$. {\bf Middle row}: Same as
the top row but for $\beta=20$. Each snapshot also shows the coefficient of
restitution, $R^2(\Pi^D, \Pi^{\theta})$ computed over the entire test set
dataset (consisting of 400 snapshots). {\bf Bottom row}: Comparison of
the power spectra, $\Pi^D(k)$ (black dashed lines) with $\Pi^{\theta}(k)$, and
the corresponding probability distributions, $P(\Pi^D)$ and $P(\Pi^{\theta})$
for each of the three cases across different values of $\beta$, shown in the
two row. Note that the colored curves can be difficult to discriminate
because of their close overlap in a consistent way with spatial error plots
such as shown  in  panels b)-d) and f)-h). The markers have been removed from k) and
l) to reduce clutter.}
\label{fig:offline}
\end{figure}
%%%%%%%%%%
We evaluate how the CNN models perform in predicting the SGS flux divergence,
$\Pi$, on the test set given the inputs, either $(\bar u, \bar v)$ or
$(\bar\omega, \bar\sigma_s, \bar\sigma_n)$. Ultimately our objective is to
construct computationally efficient neural parameterizations which lead
to accurate and stable online solutions. However, it is helpful to
examine offline test set performance as a precursor to evaluating online
performance and to potentially foreshadow a relationship between the two.
Furthermore, the presence of a ground truth, absent in online solutions where
the CNN-LES and trajectories diverge due to dynamical chaos, allows an
examination of the nature of the modeling errors.

Given our large hyperparameter sweep in the $(lr, wd)$-space, we choose to show
offline model comparisons for hyperparameter choices that lead to the best
performance on online metrics defined in Sec.~\ref{sec:diag}, as detailed
further in Sec.~\ref{sec:results}. The results are shown in
Fig.~\ref{fig:offline} where we compare individual snapshots of $\Pi^D$ from
the test set and the CNN prediction error $(\Pi^D - \Pi^{\theta})$ for
different inputs, depending on whether or not learning rate annealing was used.
The average test set $R^2$, as indicated on the corresponding snapshots in
Fig.~\ref{fig:offline}, ranges from 0.73 to 0.8 but the
prediction error shows surprisingly similar small-scale structure across
different models and learning rate modalities for both the $\beta=0$ and
$\beta=20$. This close similarity of the different models can be observed both
in the power spectra and probability distribution functions, relative to that
of the FDNS test data (bottom row of Fig.~\ref{fig:offline}). From the spectra
comparisons, we note that the disagreement between the model predictions and
data are limited primarily to the three largest wavenumbers, at the end of the spectrum.  We found that these high-wavenumber signals can not be modeled even with deep 10-layer NNs and our choice of small parameter-CNNs are not the
reason. The presence of these high-wavenumber components is a consequence of
a leakage of the spectral cutoff operator that is not removed by the Gaussian
filter. 

\subsection{Online prediction}
\subsubsection{Forecast accuracy}\label{sec:forecast}
\begin{figure}[!h]
\centering
\includegraphics[width=1\linewidth, height=.4\linewidth]{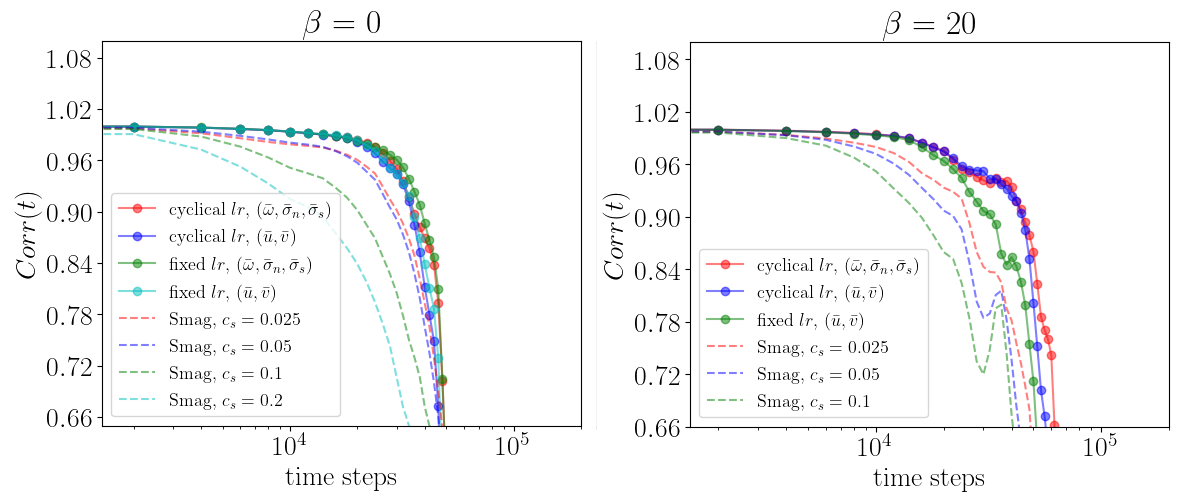} %
\hfill
  \caption{Temporal evolution of the correlation coefficient between the FDNS and multiple
	configurations of the CNN-LES models (each initialized with the same FDNS snapshot) as a function of time-steps of the DNS
	model (solid lines with markers); to get the corresponding number of time steps
	for the CNN-LES, divide by 16. Also shown are the correlations of the
	Smagorinsky-parameterized runs with the FDNS for different Smagorinsky
	constants as indicated. The CNN-LES cases shown here correspond to the best long-term online error based on a large hyperparameter search for the choice of inputs or learning rate choice; see Secns.~\ref{Sec_longterm} and \ref{Sec_hyperparamdep} below.}
\label{fig:forecast}
\end{figure}
%%%%%%%%%%%%%%%%%%%%%%%%%%%%%%%%%%%%%%%

We start by examining the short-term forecast accuracy of our CNN-LES models.
In order to achieve this, we run the vorticity closure equation
Eq.~\eqref{eq:vortFilt}, initialized with a {\mkr same} snapshot from the FDNS validation
dataset and solve forward in time with either the parameterized CNN model
learned from data, or the Smagorinsky/Leith parameterizations (we refer to all
these runs generically as LES runs). We then compare the short term evolution
of the precomputed FDNS solution with the corresponding online CNN-LES solution
and Smagorinsky/Leith runs in turn. {\mkr Adopting this protocol, the}  longer our LES runs remain correlated in time
with the ground truth FDNS, the better we consider the forecast accuracy of the
LES to be. We use standard Pearson's correlation computed at each time between
the FDNS and the LES $\bar\omega$ snapshots, and visualize these as a function
of time represented as multiples of $\Delta t$, recalling  that $\Delta
t_{LES} = 16 \Delta t$. 

We {\mkr then} define a single metric for forecast
accuracy, the decorrelation time, as the time after which the correlation
between FDNS and LES drops below 0.96 \citep{dresdner2022}; thus a longer
decorrelation time implies a better forecast accuracy. Given that we are dealing
with chaotic turbulent flows, one can also define an \textit{intrinsic}
decorrelation time of the FDNS itself, evaluated by perturbing the initial DNS
snapshot with noise and then {\mkr measuring} how fast the flow decorrelates from the
unperturbed DNS. \cite{ross2023} found that Smagorinsky models had
close to the best forecast accuracy among all their LES models in spite of poorer
long-term online performance. Therefore, we believe it suffices to compare the
relative forecast performance of our CNN-LES runs with Smagorinsky/Leith LES
runs serving as a strong baseline, without concerning ourself with the
intrinsic decorrelation of the DNS itself.

Figure~\ref{fig:forecast} {\mkr shows the resulting forecast results.} First, we note that all four types of
CNN-LES solutions (solid curves with solid markers) corresponding to cases
shown in the offline section (Sec.~\ref{sec:offline}) have remarkably similar
correlation curves.  Furthermore, the CNN-LES solutions outperform the
Smagorinsky cases (dashed lines) substantially  taking almost twice as long to decorrelate from the FDNS runs (note that the time axis is
logarithmic). {\mkr A decrease of} the Smagorinsky constant $c_s$ {\mkr results into an improvement in} the forecast
performance. {\mkr However,} as discussed in subsequent sections, decreasing {\mkr the value of $c_s$
below $0.025$} (purple curve in Fig.~\ref{fig:forecast})
causes a small-wavenumber pile up of energy {\mkr leading eventually to a} deterioration of {\mkr the long-term
statistics compared to FDNS}. 

Similar observations hold for the case with $\beta=20$, 
{\mkr but unlike the $\beta=0$
case, the forecast accuracy of CNN-LES models trained  through cyclical $lr$ is demonstrably better
than for the CNN-LES models trained with fixed $lr$. In each case, the CNN-LES runs outperforms  the Smagorinsky and Leith runs in terms of forecasting accuracy; with Leith's results very similar to the Smagorinsky's ones (not shown).}

\subsubsection{Long-term accuracy}\label{Sec_longterm}
\begin{figure}[h]
\centering
\includegraphics[width=\linewidth]{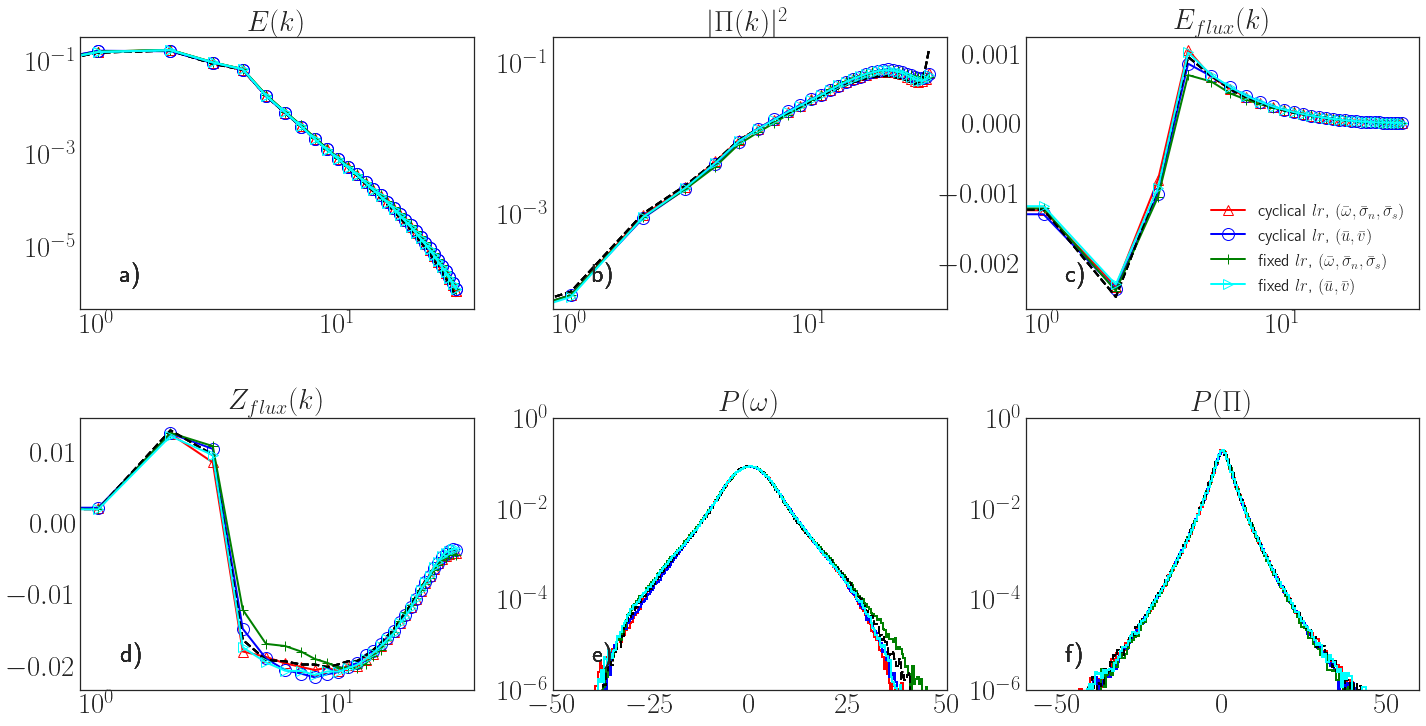} %
\hfill
	\caption{Comparison of FDNS data (dashed black line) with online runs across four different
cases shown in Fig.~\ref{fig:forecast} and Fig.~\ref{fig:offline} as indicated
on the legend, for $\beta$=0 for the diagnostic quantities indicated. {\mk Note that
the different curves are are barely discernible in a), b), f) and to a certain extent e).
The real points of difference emerge in the two cross-scale energy fluxes, c) and d);
here the two cases with $(\bar u, \bar v)$ are the most accurate. 
The markers have been removed from e) and f) to reduce clutter.}} 
%%%%%%%%%%%%%%%%%%%%%%%%%%%
\label{fig:metric_nobeta}
\end{figure}
\begin{figure}[h]
\centering
\includegraphics[width=\linewidth]{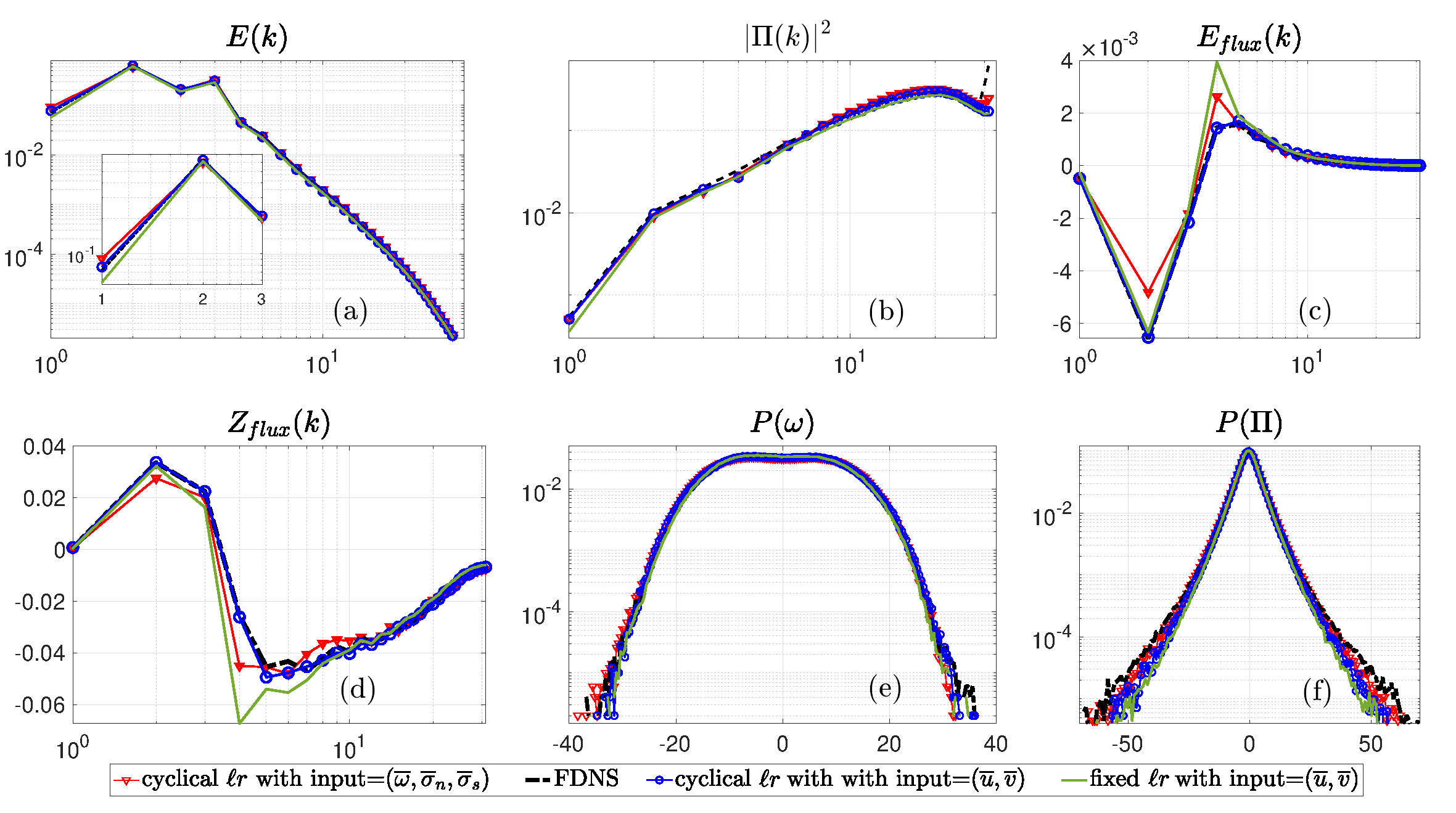} %
\hfill
	\caption{Same as Fig.~\ref{fig:metric_nobeta} but for $\beta$=20.
The case corresponding to a fixed $lr$ and $(\bar\omega, \bar\sigma_s,
\bar\sigma_n)$ inputs is not shown because of poor accuracy. \mk The inset in a)
shows a blow-up of the low-wavenumber energy spectrum of each CNN-LES model. Note that
unlike in Fig.~\ref{fig:metric_nobeta}, the CNN-LES model learnt through cyclic $lr$ with $(\bar u, \bar v)$ inputs (blue curve with circles) substantially outperforms in accuracy the other CNN-LES models.}
\label{fig:metric_beta}
\end{figure}
\begin{figure}[h]
\centering
\includegraphics[width=\linewidth]{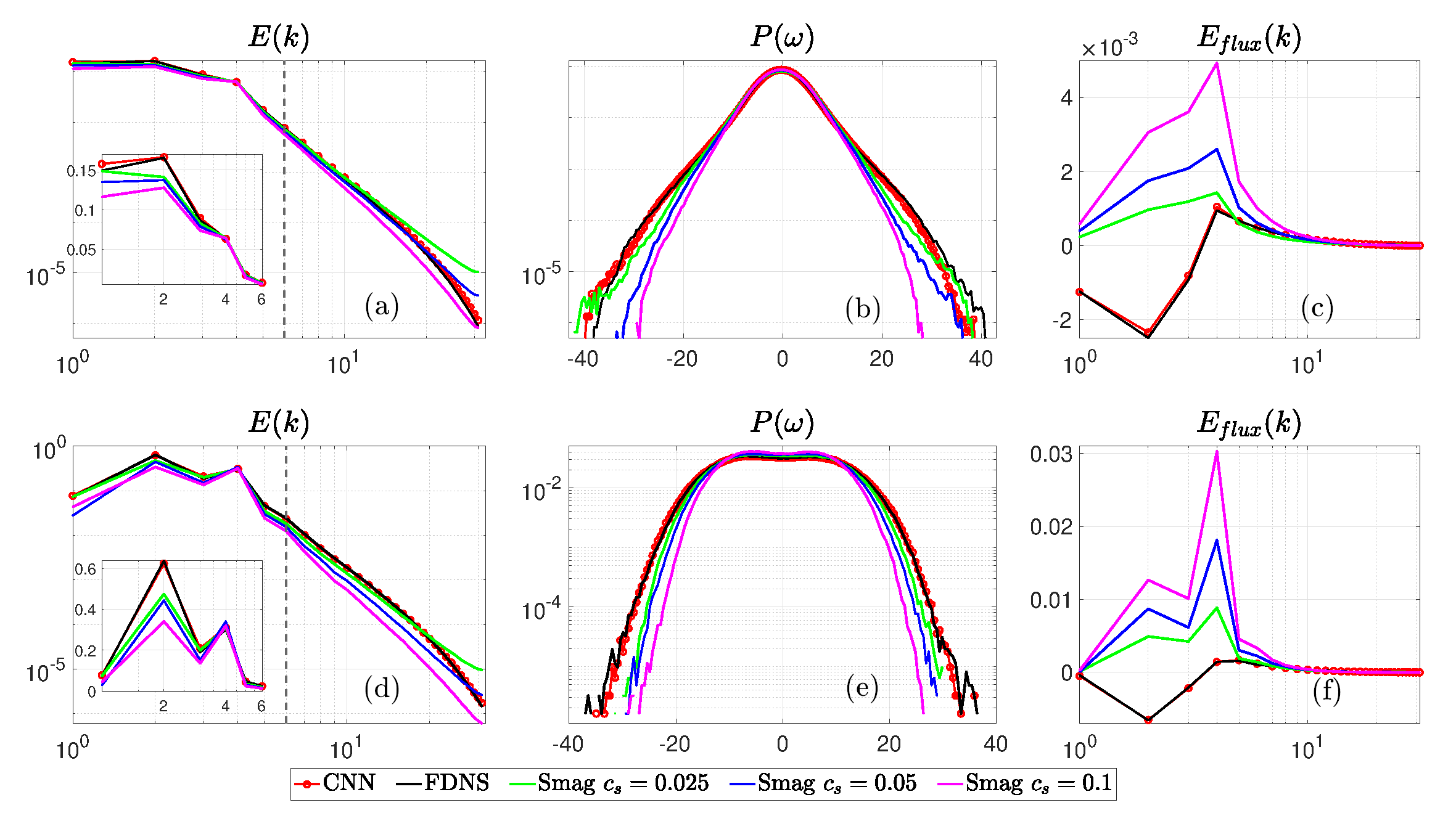} 
\hfill
	\caption{
Comparison of long-term online performance of the best CNN models (both cyclic
$lr$ cases) from Fig.~\ref{fig:metric_nobeta} and Fig.~\ref{fig:metric_beta}
with the respective Smagorinsky runs which are run for identical times as the
CNN-LES cases. Only a few of the key metrics from the earlier figures are shown
for brevity. Insets highlight the large-wavenumber comparisons between CNN-LES
and Smagorinsky spectra; note that these are shown with a linear $y$-axis
instead of logarithmic to highlight the differences further. The vertical dashed line
in (a) and (d) shows the rightmost extent of the inset. Note that, compared to the
online CNN-LES solutions, the Smagorinsky ones fail to capture the inverse
cascade.
}
\label{fig:metric_smag}
\end{figure}
%
%%%%%%%%%%%%%%%%%%%%%%%%%%%%%%%%%%%%%
\begin{figure}[!h]
\centering
\includegraphics[width=\linewidth]{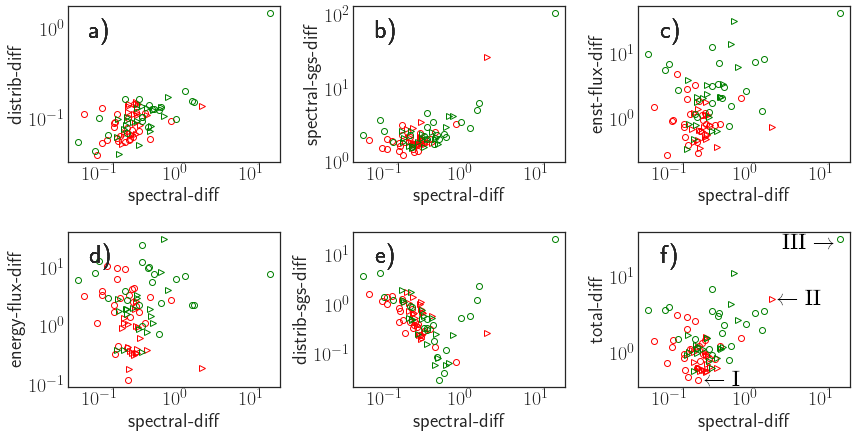} %
\hfill
	\caption{ 
		A comparison of five of the difference metrics defined in
Sec.~\ref{sec:diag} against the sixth {\mkr online} metric, the \texttt{\small spectral-diff} that measures
the difference between the energy spectra for online runs based on CNNs trained
across a range {\mkr $(lr, wd)$-values in the} hyperparameters  space. Here $\beta$=0. The red
markers denote {\mkr online} CNN-LES models trained with cyclical $lr$ while green ones with
fixed $lr$. The circles use $(\bar \omega, \bar \sigma_n, \bar \sigma_s)$ as
inputs while the triangles use $(\bar u, \bar v)$. The \texttt{\small total-diff} metric is
defined in {\mk Eq.~\eqref{eq:total}}. The cases I, II and III marked (f) are
visualized in Fig.~\ref{fig:patterns}. 
} 
\label{fig:sim}
\end{figure}
%%%%%%%%%%%%%%%%%%%%%%%%%%%%%%%%%%%%%%%%%%%%%%%

%
\begin{figure}[!h]
\centering
\includegraphics[width=\linewidth]{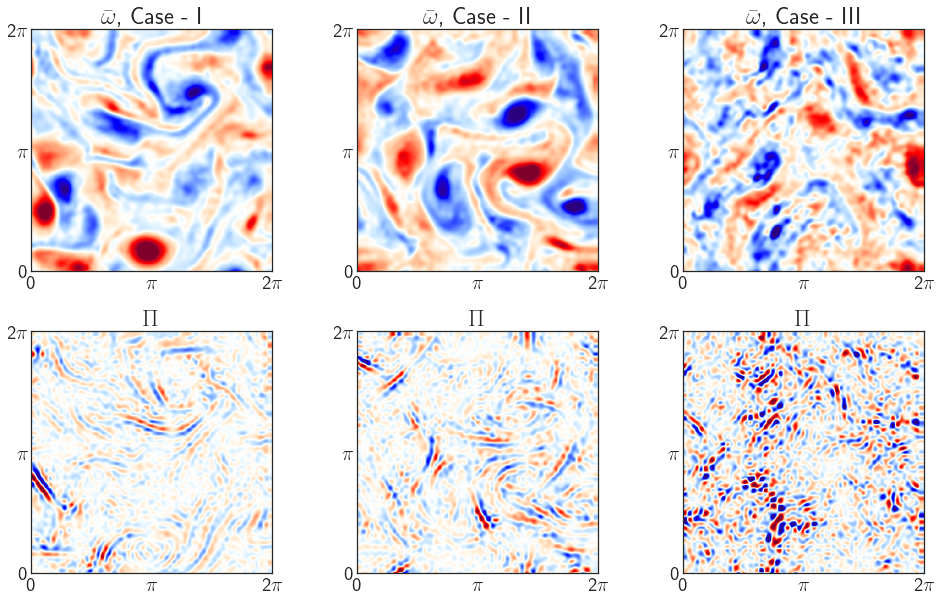} %
\hfill
	\caption{
Snapshots of $\omega$ and corresponding $\Pi$ for three online corresponding to cases marked I, II and III
found in Fig.~\ref{fig:sim}f. 
} 
\label{fig:patterns}
\end{figure}
Short term forecasts are of direct interest to problems like weather
forecasting but long term forecasts are more relevant for climate studies.
There is no \textit{a priori} reason to expect that high accuracy in the former
implies the same for the latter, especially when concerning data-driven models
whose training can be highly task specific.  In fact, \cite{dresdner2022} find
that their turbulence parameterized solutions are matched in forecast accuracy
by a purely data-driven auto-regressive NN model {\mk through} the
Encoder-Process-Decoder framework used in \citep{stachenfeld2021}, a model that
is ultimately unstable over long times.  Our objective is to validate our
CNN-LES setup for numerical stability and fidelity over long time scales, with
the forecast accuracy being a mere side-effect.  Given our aggressive choices
regarding the size of the CNN we run our models for substantially longer times
than {\mk in other recent works concerned with neural turbulent closures}.
\cite{guan2022} compute their online CNN-LES runs for\footnote{Note that
our computational setup is identical to theirs, except our $Re$ value is
marginally larger.} 2$\times$10$^6$ $\Delta t$ but we compute to
$T=6\times$10$^6$ $\Delta t$ {\mk corresponding to} 375,000 $\Delta t_{LES}$. As
explained in Sec.~\ref{sec:diag}, this also ensures convergence of metrics like
the probability distribution function, which can now be computed directly
without resorting to kernel density estimation. 

In this section, we highlight the long term accuracy of the four specific CNN
models described in Sec.~\ref{sec:offline} and Sec.~\ref{sec:forecast} for the
six quantities chosen in Sec.~\ref{sec:diag} as diagnostics for measuring
solution fidelity; these are the 1-dimensional time-averaged energy spectrum, $E(k)$, the
SGS divergence spectrum, $|\Pi(k)|^2$, the cross-scale
energy and enstrophy fluxes, $E_{flux}(k)$ and $Z_{flux}(k)$ and the 1-D
probability distribution functions, $P(\bar\omega)$ and $P(\Pi)$ computed over
the course of the entire online run. A comparison of the FDNS runs with the
corresponding four CNN-LES runs is shown in Fig.~\ref{fig:metric_nobeta}. We
find that all four cases have good agreement across the six shown metrics, with
the highest overall accuracy being observed in the two cases with cyclical
learning rate but different inputs. The SGS spectrum, $|\Pi(k)|^2$ and PDF,
$P(\Pi)$ has as high a degree of accuracy in the online runs as in the offline
tests (Fig.~\ref{fig:offline}) which also ties in with the high accuracy of the
obtained structural flow metrics like $E(k)$ and $P(\omega)$ and the dynamical
metrics of the cross-scale fluxes. An implication of this result is that
Galilean invariance is, evidently, not a particular difficult physical symmetry
to learn in these cases. 

However, these results do not translate to the case with $\beta=20$
(Fig.~\ref{fig:metric_beta}) when the run corresponding to cyclical annealed
$lr$ and $(\bar u, \bar v)$ input {\mk (blue curves in
Fig.~\ref{fig:metric_beta})} substantially outperforms on all metrics except
for $P(\Pi)$. While the role of the cyclical annealing in model robustness and
fidelity is broadly expected, it is surprising that choosing Galilean
invariance \textit{a priori} in our modeling (through using  $(\bar \omega,
\bar \sigma_n, \bar \sigma_s)$ as inputs) actually \textit{hinders} online
accuracy. The precise reason  is unclear but it evidently has to
do with the presence of strong eddy-driven zonal jets in the $\beta=20$ case
that are absent when $\beta=0$. This observation seems to imply that for the case of
geophysical turbulence, it might be better to choose  $(\bar u, \bar v)$ as
inputs to the CNNs.  It is also interesting to note that while $\Pi(k)$ is
accurately predicted as is the vorticity distribution, $P(\bar\omega)$,
the tails of the distribution $P(\Pi)$ are missed.

Interestingly, we observe that our CNN solutions {\mkr have here} high online accuracy
across both short and long-time scales (i.e.~across both ``weather'' and
``climatic'' regimes).  Previous results do not report on such solutions. 
 While \cite{dresdner2022} do not examine long-time fidelity,
\cite{frezat2022} find that their best long term accurate solutions (trained
using their fully online methodology) actually have poor forecast accuracy and
lose out to an offline-trained CNN model that is numerically unstable at long
times. Consistent with this result, \cite{ross2023} find that their forecast
accuracy does not in general correlate with long-term fidelity. To underscore
this result further, we evaluate the online performance of the best CNN-LES
solutions for the $\beta=0$ and $\beta=20$ cases (among the cases shown in
Figs. \ref{fig:metric_nobeta} and \ref{fig:metric_beta}) in
Fig.~\ref{fig:metric_smag} with reference to various Smagorinsky cases shown
earlier in Fig.~\ref{fig:forecast}. Note that the Smagorinsky solution with the
best accuracy in capturing the distribution, $P(\bar\omega)$, {\mkr (corresponding to $c_s = 0.025$)}
has a pile up of energy at small scales (Fig.~\ref{fig:metric_smag}ab).
Furthermore, the Smagorinsky cases fail to capture the tails of the vorticity
distribution that the CNN-LES solutions do consistently; this is especially
true for the jets cases (Fig.~\ref{fig:metric_smag}f).

Our online accuracy for the jets case is comparable to that obtained by
\cite{frezat2022} using their online-learning framework.  A similar attempt by
\cite{ross2023} using a similar offline-learning approach as ours for
parameterizing their corresponding jets case was, however, unsuccessful;  
their CNN completely failing to model jets after successfully solving their
$\beta=0$ case.   \cite{ross2023}  use this result to remark on the lack of robustness
of the data-driven CNN approach. Our results are in contradistinction with theirs and we do not thus share the same conclusions.
It is unclear {\mkr though why the results of \citep{ross2023} regarding the $\beta$-case}
differ from ours. While it is plausible that their {\mkr quasi-geostrophic turbulent model} could be a more challenging test case, their Reynolds number is
actually a lot lower than ours.  The reasons behind the discrepency  between our online CNN-LES solutions and those of \citep{ross2023} may also lie in technical reasons such as 
sub-optimal hyperparameter searching, the depth of the CNNs used, and the choices of training procedures. We further discuss this issue
in the Discussion section.  

Finally, our best models, shown here and the preceding two sections are
cherry picked from a large range of models trained over a range of
hyperparameters.  In the next section we examine the behavior of our CNN-LES models across
this {\mk hyperparameter space} depending on the training methodology or input choices.

\subsubsection{Hyperparameter dependence of model fidelity}\label{Sec_hyperparamdep}
\begin{figure}[!h]
\centering
\includegraphics[width=\linewidth]{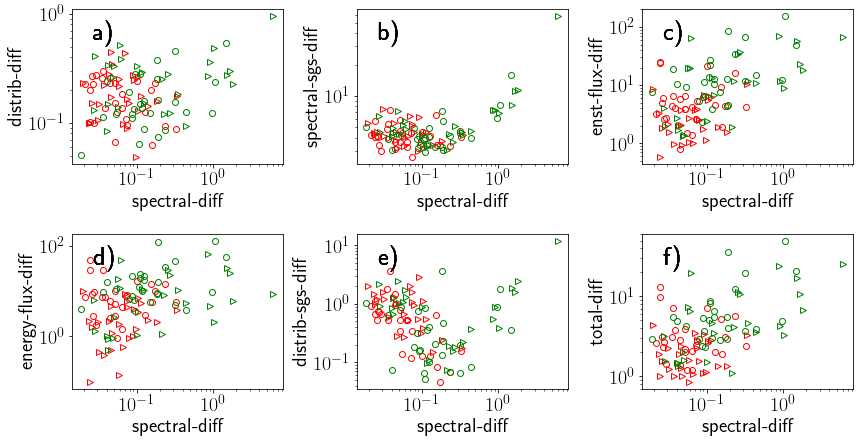} %
\hfill
	\caption{Same as Fig.~\ref{fig:sim} but for $\beta$=20. Here again one observes an overall better online performance of the CNN-LES models trained with cyclical $lr$ rather than fixed ones (red vs green markers). }
\label{fig:simbeta}
\end{figure}
%%%%%%%%%%%%%%%%%%%%%%%%%%%%
In preceding sections we have demonstrated that by searching in the $(lr, wd)$-hyperparameter space we can obtain high-fidelity solutions with shallow CNNs.
However, it is important to know how common or rare such high-fidelity
solutions might be to find, and to in turn to be able to find the most accurate
online solutions easily given such a large number of runs. Such an evaluation,
though seemingly mundane and somewhat tedious, is especially relevant for
foreshadowing the challenges in parameterizing more complex flow configurations
observed in the climate system when such large hyperparameter searches might
not be viable.

In Sec.~\ref{sec:diag}, we {\mkr define} six different difference metrics based on
the six quantities shown in Fig.~\ref{fig:metric_nobeta} and
Fig.~\ref{fig:metric_beta} that {\mkr compare} errors in the online CNN-LES solutions
relative to the FDNS solutions. {\mkr Here, we} construct a metric that {\mkr consists of
 simply summing up} the six metrics, defined thus as 
%%%%%%%%%%%%%%%%%%%
\begin{equation}
\begin{split}
\texttt{\small total-diff} = & \texttt{\small (spectral-diff + spectral-sgs-diff + enst-flux-diff} \\
		& \texttt{\small + energy-flux-dim + distrib-sgs-diff)/5}\,.
\end{split}
\label{eq:total}
\end{equation}
%%%%%%%%%%%%%%%%%%%
To understand how these metrics vary in the hyperparameter space, we display the 5
metrics, \textit{other than} \texttt{\small spectral-diff} (which measures the accuracy in
estimating the energy spectrum) relative to  \texttt{\small spectral-diff}. This helps us assess
how the various metrics {\mkr relate (implicitly) to each other} by observing
their variation {\mkr against} \texttt{\small spectral-diff}.  Figures \ref{fig:sim} and
\ref{fig:simbeta} show the results for $\beta=0$ and $\beta=20$
respectively. We examine each of these two cases as the results show
different variations of the metrics across the $(lr, wd)$-space. We start by
examining the {\mkr \texttt{\small total-diff}} metric (Fig.~\ref{fig:sim}f) which highlights two
important points. First, {\mkr we observe} the cyclical annealed cases (red markers) are
more tightly clustered and closer to the origin (i.e.~lower errors) while the
fixed $lr$ cases (green markers) have a much larger spread in performance. This {\mkr feature}
indicates that not only does the cyclic $lr$ procedure lead to more robust
solutions overall by producing a narrower range of behaviors as the
hyperparameters are changed, they are also overall more accurate as well. {\mkr Such properties are}  even more pronounced for $\beta=20$ (Fig.~\ref{fig:simbeta}f) when
the lower accuracy and robustness of the annealed $lr$ cases stand out even
more starkly. Second, within the cyclic $lr$ cases (red markers), solutions
that use $(\bar \omega, \bar \sigma_n, \bar \sigma_s)$ as inputs (red circles)
perform best when $\beta=0$ though not by a lot.  However, when $\beta=20$
almost all the best solutions are for the cases which use $(\bar u, \bar v)$ as
inputs (red triangles).  The cyclic $lr$ cases also have substantially lower
flux errors (Fig.~\ref{fig:simbeta}cd) compared to the fixed $lr$ cases though
the other metrics are harder to distinguish. Curiously for both $\beta=0$ and
$\beta=20$, the fidelity of $P(\Pi)$ (Fig.~\ref{fig:sim}e and
Fig.~\ref{fig:simbeta}e) seems to have the opposite relationship with the flow
metrics with the fixed $lr$ cases having higher accuracy in this metric. We
believe this is not a significant issue as tail errors in $P(\Pi)$ are not
insignificant even for the offline cases ({\mkr Figs.~\ref{fig:offline}k and \ref{fig:offline}l}).  The
cyclic $lr$ cases, however, outperform on the  \texttt{\small spectral-sgs-diff} metric
(Fig.~\ref{fig:sim}e and Fig.~\ref{fig:simbeta}e).  

 Our exhaustive parameter search allows us to highlight a specific case with extremely large
online error, an order of magnitude larger than that in the best case, {\mkr while being}
numerically stable for long times (marked as Case III in
Fig.~\ref{fig:sim}f).  To understand this anomalous solution better, we
visualize snapshots of $\bar\omega$ and $\Pi$ for three cases corresponding to
the CNN-LES runs with best  {\mkr and middling online runs along with this} worst online {\mkr run}.  A surprising fact
is that both I (the best case) and II {\mkr (the ``middle'' case)} have very similar flow structure while
III looks noisy and unphysical. Curiously though, the three cases have similar
offline accuracy, again highlighting the challenges in predicting online
accuracy from from offline accuracy.

The findings above also raise questions about the relationship between the
large number of online-stable CNN parameterizations that we find as part of our
hyperparameter searches. The close structural similarity of the SGS error observed
in Fig.~\ref{fig:offline} across different input choices and optimization methods indicates that we may be learning a common family of
parameterizations. While one might hope to find a single unique
parameterization that is globally optimal under some accuracy metrics for
complex turbulent flows, this can be challenging, if at all possible, to infer
from finite data sets. One might wonder how these results change for a fixed
neural network architecture but in the infinite data limit. Such questions are tied to the study of ergodicity properties. If ergodicity applies, the uniqueness (and existence) of a parameterization that averages out optimally the small scales is guaranteed in the infinite data limit  \cite[Theorem 4]{CLM19_closure} although the use of finitely sampled data can lead to various admissible solutions with different scoring as reported in Fig.~\ref{fig:sim} and Fig.~\ref{fig:simbeta}. A detailed study of this phenomenon under the lens of a possible ergodicity of the underlying forced flows \citep{FMRT01,CGH12,hairer2006ergodicity,hairer2011theory} goes beyond the scope of the present work.

\subsubsection{Offline vs online predictions}
\begin{figure}[!h]
\centering
\includegraphics[width=\linewidth]{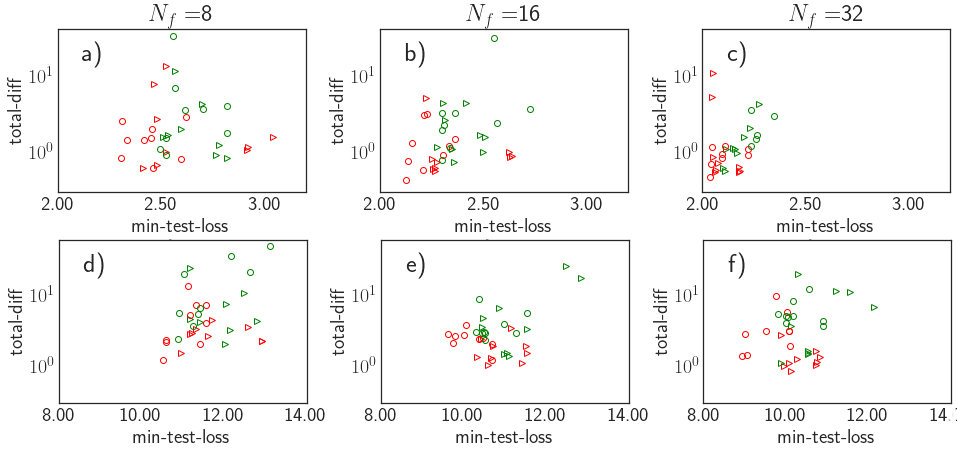} %
\hfill
	\caption{Comparison of online {\mkr \texttt{\small total-diff}} metric vs \texttt{\small offline test-error}
over the $(lr, wd)$-hyperparameter space for a)-c) $\beta=0$ and d)-f)
$\beta=20$. The results are grouped by the different values of the convolutional
filters used, given by $N_f$; details of the CNN architecture are given in Sec.~\ref{subsec:arch}. {\mkr Markers color-coding is the same as in Fig.~\ref{fig:simbeta}.}} 
\label{fig:onoff}
\end{figure}
%
%%%%%%%%%%%%%

The ultimate goal of any data-driven parameterization {\mkr is to reach high-quality} online fidelity
and accuracy.  {\mkr To achieve such a feat requires typically to} run the model over long times
and evaluating {\mkr the online solutions across} various metrics {\mkr such as reported the sections above}. 
{\mkr Such an approach becomes particularly expensive as one tests  over a} wide swaths in the hyperparameter space even when using {\mkr shallow neural networks like in this study}. Ideally one would hope that one could simply compute
offline accuracy on the held-out test set and use that to estimate online
accuracy. {\mkr But this remains idealistic. In that respect, Figure \ref{fig:onoff} compares} offline accuracy {\mkr with} the \texttt{\small total-diff}
online accuracy measure defined in \eqref{eq:total} across the $(lr, wd)$-hyperparameter space. The results are shown for $\beta=0$ in
Fig.~\ref{fig:onoff}a-c and for $\beta=20$ in Fig.~\ref{fig:onoff}d-f.
Unlike in the previous section, we separate the results for different filter
sizes of the hidden layers. It {\mkr can be observed} that as the number $N_f$  of convolutional
filters  {\mkr is} increased corresponding to an increase in the number of CNN
parameters, the offline error (the $x$-axis) systematically decreases with
increasing $N_f$, {\mkr as manifested by} the shift of the cloud of points towards
the origin of the $x$-axis. However, the corresponding decrease in online error
(the $y$-axis) is actually rather small and consequently difficult to discern.
This {\mkr observation} highlights the crucial point that a large decrease in offline error might
lead to only small gains in online CNN-LES performance. {\mkr It} also foreshadows
the result that deeper CNNs that would inevitably improve offline accuracy do
not necessarily lead to more accurate online solutions.

Concurrently, we note that the correlation between online and offline
performance remains weak implying that diagnosing the latter is insufficient
for knowledge of the former, our actual objective. One reason is that our
metrics for offline performance, the mean square error, calculated over 400
snapshots and spatially averaged {\mkr through the norm involved therein,} does not reflect {\mkr necessarily} the dynamical differences  in the data produced by the different CNN-LES models.
{\mkr It is} conceivable that more
physically grounded metrics of offline performance might in fact improve
online skills without running more expensive CNN-LES runs
for long times; this remains an area of active investigation. 

\section{Discussion and concluding remarks}
%%%%%%%%%%%%%%%%%%%%%%%%%%
\subsection{Physical and computational implications of SGS near-locality} 
 The primary result in this work is the {\mk effective} learning of accurate
parameterizations of two-dimensional and geostrophic turbulent flows with
shallow two-layer CNNs. This {\mk accomplishment essentially implies that the SGS
stresses for these problems} have a spatially {\mkr nearly} local (or weakly non-local)
dependence on the coarse fields. In other words we have inferred the physics of
this class of problems through direct construction of a spatial {\mkr nearly} local
parameterization. This is not in general true for all classes turbulent flows;
for example convective motions in the the oceanic surface-mixed layer and
broadly in the Earth's troposphere (processes that also result in cloud
formation and subsequent precipitation) can be  strongly correlated across a
significant fraction of the vertical extent within which they take place. Recent ML
models that successfully parameterize vertical fluxes in the atmosphere use NNs
that span the most of the air column \citep{yuval2020, yuval2021}. Similarly,
traditional empirical physics-based parameterizations in the oceanic surface
mixed layer employ a simple vertically non-local flux to represent convective
turbulence. While current class of atmospheric ML-based parameterizations do
not represent the horizontal eddy-fluxes, evidently for reasons of simplicty,
our work here suggests that more complete representations would likely be
spatially (nearly) local in the horizontal and have a greater degree of non-locality in
the vertical; in the ocean the non-locality should be limited to the extent of
the surface-mixed layer where air-sea fluxes are actively felt. 
%%%%%%%%%%%%%%%%%%%%%%%%%%%%%%%%%

A secondary implication of this work is for purposes of numerical computation
of oceanic, atmospheric and climate models. These models often rely on domain
decomposition in the horizontal for numerical solution over large clusters of
compute nodes, with inter-process communications typically limited to a small
number of points on the boundaries of the sub-domains. Thus employing
data-driven parameterizations precludes a high degree of spatial non-locality
which would make the inter-process exchanges prohibitively expensive; note that
this is not an issue in the vertical direction (i.e.~normal to the Earth's
motions). Our results above demonstrate that the horizontal SGS fluxes can be
modeled through shallow CNN models and are thus easier to incorporate into existing
pipelines than by using more cumbersome deep CNNs. 

%%%%%%%%%%%%%%%%%%%%%%%%%%%%%%%%%%%%%%
\subsection{Theoretical consequences and towards interpretability.} 
Our neural closure results with shallow CNNs presented in this study are valid
for cutoffs within the inertial range and for high Reynolds numbers. This
problem is known to be difficult as small errors at the level of the SGS
typically amplify the errors at the large scales due to the inverse cascade
\citep{piomelli1991subgrid,jansen2014parameterizing}. To dispose of SGS
parameterizations  at low cutoff levels for such turbulent flows with a
controlled error is thus one of the challenges to resolve.  The accuracy and
stability of our closure results are thus strongly supportive for the existence
of a nonlinear function $\Phi_{CNN}$ such that the SGS, $\Pi$, satisfies, after
spin up, a relation of the form
\beq\label{Eq_fundamental_relation}
\Pi=\Pi_{CNN}(\overline{u},\overline{v})+\epsilon,
\eeq
where the residual $\epsilon$ is a spatio-temporal function whose fluctuations
are controlled and small in a mean square sense. In
Eq.~\eqref{Eq_fundamental_relation}, $\Pi_{CNN}$ denotes the function found by
means of shallow CNNs trained by minimizing our loss function in (v). 
Actually, \eqref{Eq_fundamental_relation} is a consequence  of the very
construction of $\Pi_{CNN}$ obtained by minimization of our loss function
\eqref{eq:loss} along with its regularization terms.

%%%%%%%%%%%%%%%%%%%%%%%%%%%%%%%
 We have shown that with the appropriate hyperparameter searching in the
learning rate and weight decay coefficient space, as well as usage of  cyclical
learning rate annealing, small residuals can be reached offline---with shallow
and weakly local CNNs---while leading to stable and accurate online solutions.
As  $\Pi_{CNN}$ is a nonlinear functional of the coarse-grained variables only,
our good closure skills  suggest that the knowledge of $\Pi_{CNN}$ is amply
sufficient to achieve good performances in particular in the $\beta$-case when
compared to other recent neural closures \citep{ross2023}.  

As such, our neural closure results as those of
\citep{maulik2019,kochkov2021,zanna2020,subel2022explaining} rule out  for
turbulent problems, even at low cutoffs,  the use of memory  terms in the
Mori-Zwanzig (MZ) interpretation \citep{GKS04,Chorin_Hald-book}; memory terms
that have been thus unecessarily praised in other closure studies relying on
the MZ formalism; see
e.g.~\citep{miyanawala2017efficient,parish2017non,ma2019model}.

To the contrary, a good approximation of the conditional expectation, namely
the best nonlinear functional averaging out the unresolved variables as
conditioned on the coarse variables, is sufficient for the closure of forced
two-dimensional turbulence problems at high $Re$.  In this study, $\Pi_{CNN}$
with a small $\epsilon$ provides such an approximation and as such is likely to
relate to the existence of an underlying optimal parameterizing manifold (OPM)
linking the small and large scales in a least squares sense 
\citep{chekroun2017emergence,CLM19_closure,chekroun2021stochastic} as predicted
by the theory of OPMs \cite[Theorem 5]{CLM19_closure}.

Our neural turbulent closure results together with related recent studies
restore thus some credentials to ideas proposed in the late 80s by
\cite{FMT88,foias1991approximate} envisioning two-dimensional turbulence as
essentially finite-dimensional with turbulent solutions lying in some thin
neighborhood, in a mean square sense, of a finite-dimensional manifold; see also 
\citep[Eq.~(1.5)]{CLM19_closure}. These  ideas were  watered down as shown to
be valid only for cutoff wave numbers within or close to the dissipation range
\citep{pascal1992nonlinear} when relying on {\it traditional analytic
parameterizations} such as initially proposed in \citep{FMT88}.  The usage of
neural networks invites us thus to revise such conclusions based on a limited
class of analytic formulas, and  sheds actually new lights on this old problem
as pushing the validity of relationships such as
\eqref{Eq_fundamental_relation} for cutoff within the inertial range.  The
discovery of nearly-local shallow CNN-parameterizations to achieve this feat is
likely to be interpretable and generalizable because of its intrinsic low
dimensionality.  We hope thus to reconcile the previous failures in analytic
attempts  with the recent empirical successes by seeking for new analytic
formulas for closure that would exploit the discovery of our nearly-local
shallow CNNs.

%%%%%%%%%%%%%%%%%%%%%%%%%%%%%%%%
\subsection{Machine Learning vs physical design choices} 
In their recent work, \cite{ross2023} suggest that the focus of ML-based
parameterizations should be less on ML details like NN architectures or
optimization techniques but should instead be focused on physical design
choices, in this case on the choice of the inputs/outputs or the
coarse-graining filter. Based on our results, we suggest that one should in fact focus on both
aspects, and in some cases these are closely related e.g.~our choice of the
2-layer CNN architecture based on the {\mkr nearly local character} of the SGS  {\mkr coarse-field components interacting with the small scales}. {\mkr This study shows} that {\mkr the choice of} optimization techniques like cyclical
annealed $lr$ can make the learning task substantially easier and generally
more robust; this would be even more relevant when the problems being
considered {\mkr are not} simplified {\mkr turbulent models}  like the ones {\mkr dealt with} here and in
the studies referred to as in this study. Furthermore, hyperparameter
searches, while tedious, can lead to substantial improvements in task accuracy
and efficiency, and therefore should not be ignored.

%%%%%%%%%%%%%%%%%%%%%%%%%%%%%%%%%%
\section*{Acknowledgments}
This work is supported by the Office of Naval Research (ONR) Multidisciplinary University Research Initiative (MURI) grant N00014-20-1-2023. This work is also partially supported (MDC) by the European Research Council (ERC) under the European Union's Horizon 2020 research and innovation program [Grant Agreement No. 810370].
%%%%%%%%%%%%%%%%%%%%%%%%%%%%%%%%%%%%%%%%%%%

\bibliographystyle{ametsoc2014AG}
%%%%%%%%%%%%%%%%%%%%%%%%%%%%%%%%%%%%%%%%%%%%%%
%\bibliographystyle{aipnum4-1}
%\bibliographystyle{apsrev}

%\bibliographystyle{plain}
%\bibliography{references_MC}

\begin{thebibliography}{67}
\providecommand{\natexlab}[1]{#1}
\providecommand{\url}[1]{\texttt{#1}}
\renewcommand{\UrlFont}{\rmfamily}
\providecommand{\urlprefix}{URL }
\expandafter\ifx\csname urlstyle\endcsname\relax
  \providecommand{\doi}[1]{doi:\discretionary{}{}{}#1}\else
  \providecommand{\doi}{doi:\discretionary{}{}{}\begingroup
  \urlstyle{rm}\Url}\fi
\providecommand{\eprint}[2][]{\url{#2}}

\bibitem[{Ba et~al.(2016)Ba, Kiros,, and Hinton}]{ba2016}
Ba, J.~L., J.~R. Kiros, and G.~E. Hinton, 2016: Layer normalization.
  \textit{arXiv preprint arXiv:1607.06450}.

\bibitem[{Bachman et~al.(2017)Bachman, Fox-Kemper,, and
  Pearson}]{bachman2017Leith}
Bachman, S.~D., B.~Fox-Kemper, and B.~Pearson, 2017: A scale-aware subgrid
  model for quasi-geostrophic turbulence. \textit{Journal of Geophysical
  Research: Oceans}, \textbf{122~(2)}, 1529--1554.

\bibitem[{Bakas et~al.(2015)Bakas, Constantinou,, and Ioannou}]{bakas2015}
Bakas, N.~A., N.~C. Constantinou, and P.~J. Ioannou, 2015: S3t stability of the
  homogeneous state of barotropic beta-plane turbulence. \textit{Journal of the
  Atmospheric Sciences}, \textbf{72~(5)}, 1689--1712.

\bibitem[{Bengio and Delalleau(2011)Bengio, and Delalleau}]{bengio2011}
Bengio, Y., and O.~Delalleau, 2011: On the expressive power of deep
  architectures. \textit{Algorithmic Learning Theory: 22nd International
  Conference, ALT 2011, Espoo, Finland, October 5-7, 2011. Proceedings 22},
  Springer, 18--36.

\bibitem[{Botev et~al.(2010)Botev, Grotowski,, and Kroese}]{Botev_al10}
Botev, Z.~I., J.~F. Grotowski, and D.~P. Kroese, 2010: Kernel density
  estimation via diffusion. \textit{The Annals of Statistics}, \textbf{38~(5)},
  2916--2957.

\bibitem[{Brachet et~al.(1988)Brachet, Meneguzzi, Politano,, and
  Sulem}]{brachet1988}
Brachet, M., M.~Meneguzzi, H.~Politano, and P.~Sulem, 1988: The dynamics of
  freely decaying two-dimensional turbulence. \textit{Journal of Fluid
  Mechanics}, \textbf{194}, 333--349.

\bibitem[{Brenowitz et~al.(2020)Brenowitz, Henn, McGibbon, Clark, Kwa, Perkins,
  Watt-Meyer,, and Bretherton}]{brenowitz2020}
Brenowitz, N.~D., B.~Henn, J.~McGibbon, S.~K. Clark, A.~Kwa, W.~A. Perkins,
  O.~Watt-Meyer, and C.~S. Bretherton, 2020: Machine learning climate model
  dynamics: Offline versus online performance. \textit{arXiv preprint
  arXiv:2011.03081}.

\bibitem[{Carnevale et~al.(1991)Carnevale, McWilliams, Pomeau, Weiss,, and
  Young}]{carnevale1991}
Carnevale, G., J.~McWilliams, Y.~Pomeau, J.~Weiss, and W.~Young, 1991:
  Evolution of vortex statistics in two-dimensional turbulence.
  \textit{Physical review letters}, \textbf{66~(21)}, 2735.

\bibitem[{Carnevale et~al.(1992)Carnevale, McWilliams, Pomeau, Weiss,, and
  Young}]{carnevale1992}
Carnevale, G., J.~McWilliams, Y.~Pomeau, J.~Weiss, and W.~Young, 1992: Rates,
  pathways, and end states of nonlinear evolution in decaying two-dimensional
  turbulence: Scaling theory versus selective decay. \textit{Physics of Fluids
  A: Fluid Dynamics}, \textbf{4~(6)}, 1314--1316.

\bibitem[{Chekroun et~al.(2021)Chekroun, Liu,, and
  McWilliams}]{chekroun2021stochastic}
Chekroun, M., H.~Liu, and J.~McWilliams, 2021: Stochastic rectification of fast
  oscillations on slow manifold closures. \textit{Proc. Natl. Acad. Sci. USA},
  \textbf{118~(48)}, e2113650\,118, \doi{10.1073/pnas.2113650118}.

\bibitem[{Chekroun and Glatt-Holtz(2012)Chekroun, and Glatt-Holtz}]{CGH12}
Chekroun, M.~D., and N.~E. Glatt-Holtz, 2012: {Invariant measures for
  dissipative dynamical systems: Abstract results and applications}.
  \textit{Commun. Math. Phys.}, \textbf{316}, 723--761,
  \doi{10.1007/s00220-012-1515-y}.

\bibitem[{Chekroun et~al.(2020)Chekroun, Liu,, and McWilliams}]{CLM19_closure}
Chekroun, M.~D., H.~Liu, and J.~McWilliams, 2020: {Variational approach to
  closure of nonlinear dynamical systems: Autonomous case}. \textit{Journal of
  Statistical Physics}, \textbf{179}, 1073--1160,
  \doi{10.1007/s10955-019-02458-2}.

\bibitem[{Chekroun et~al.(2017)Chekroun, Liu,, and
  McWilliams}]{chekroun2017emergence}
Chekroun, M.~D., H.~Liu, and J.~C. McWilliams, 2017: The emergence of fast
  oscillations in a reduced primitive equation model and its implications for
  closure theories. \textit{Computers \& Fluids}, \textbf{151}, 3--22.

\bibitem[{Chorin and Hald(2006)Chorin, and Hald}]{Chorin_Hald-book}
Chorin, A., and O.~Hald, 2006: \textit{{Stochastic Tools in Mathematics and
  Science}}. No. 147, {Surveys and Tutorials in the Applied Mathematical
  Sciences}, Springer New York.

\bibitem[{Dresdner et~al.(2022)Dresdner, Kochkov, Norgaard,
  Zepeda-N{\'u}{\~n}ez, Smith, Brenner,, and Hoyer}]{dresdner2022}
Dresdner, G., D.~Kochkov, P.~Norgaard, L.~Zepeda-N{\'u}{\~n}ez, J.~A. Smith,
  M.~P. Brenner, and S.~Hoyer, 2022: Learning to correct spectral methods for
  simulating turbulent flows. \textit{arXiv preprint arXiv:2207.00556}.

\bibitem[{Eyink(2005)}]{eyink2005}
Eyink, G.~L., 2005: Locality of turbulent cascades. \textit{Physica D:
  Nonlinear Phenomena}, \textbf{207~(1-2)}, 91--116.

\bibitem[{Eyink and Aluie(2009)Eyink, and Aluie}]{eyink2009}
Eyink, G.~L., and H.~Aluie, 2009: Localness of energy cascade in hydrodynamic
  turbulence. {I}. {S}mooth coarse graining. \textit{Phys. of Fluids},
  \textbf{21~(11)}, 115\,107.

\bibitem[{Farrell and Ioannou(2007)Farrell, and Ioannou}]{farrell2007}
Farrell, B.~F., and P.~J. Ioannou, 2007: Structure and spacing of jets in
  barotropic turbulence. \textit{Journal of the atmospheric sciences},
  \textbf{64~(10)}, 3652--3665.

\bibitem[{Fisher(2008)}]{fisher2008}
Fisher, B., 2008: The cross-correlation and wiener-khinchin theorems.
  \textit{Journal of Neuroscience}, 8107--8115.

\bibitem[{Foias et~al.(2001)Foias, Manley, Rosa,, and Temam}]{FMRT01}
Foias, C., O.~Manley, R.~Rosa, and R.~Temam, 2001: \textit{{Navier-Stokes
  Equations and Turbulence}}, Vol.~83. Encyclopedia of Mathematics and its
  Applications. Cambridge University Press.

\bibitem[{Foias et~al.(1988)Foias, Manley,, and Temam}]{FMT88}
Foias, C., O.~Manley, and R.~Temam, 1988: Modeling of the interaction of small
  and large eddies in two-dimensional turbulent flows. \textit{RAIRO Mod\'el.
  Math. Anal. Num\'er.}, \textbf{22~(1)}, 93--118.

\bibitem[{Foias et~al.(1991)Foias, Manley,, and Temam}]{foias1991approximate}
Foias, C., O.~P. Manley, and R.~Temam, 1991: Approximate inertial manifolds and
  effective viscosity in turbulent flows. \textit{Physics of Fluids A: Fluid
  Dyn.}, \textbf{3~(5)}, 898--911.

\bibitem[{Frezat et~al.(2022)Frezat, Sommer, Fablet, Balarac,, and
  Lguensat}]{frezat2022}
Frezat, H., J.~L. Sommer, R.~Fablet, G.~Balarac, and R.~Lguensat, 2022: A
  posteriori learning for quasi-geostrophic turbulence parametrization.
  \textit{arXiv preprint arXiv:2204.03911}.

\bibitem[{Givon et~al.(2004)Givon, Kupferman,, and Stuart}]{GKS04}
Givon, D., R.~Kupferman, and A.~Stuart, 2004: Extracting macroscopic dynamics:
  model problems and algorithms. \textit{Nonlinearity}, \textbf{17~(6)},
  R55--R127.

\bibitem[{Guan et~al.(2022{\natexlab{a}})Guan, Chattopadhyay, Subel,, and
  Hassanzadeh}]{guan2022stable}
Guan, Y., A.~Chattopadhyay, A.~Subel, and P.~Hassanzadeh, 2022{\natexlab{a}}:
  {Stable a posteriori LES of 2D turbulence using convolutional neural
  networks: Backscattering analysis and generalization to higher Re via
  transfer learning}. \textit{Journal of Computational Physics}, \textbf{458},
  111\,090.

\bibitem[{Guan et~al.(2022{\natexlab{b}})Guan, Subel, Chattopadhyay,, and
  Hassanzadeh}]{guan2022}
Guan, Y., A.~Subel, A.~Chattopadhyay, and P.~Hassanzadeh, 2022{\natexlab{b}}:
  {Learning physics-constrained subgrid-scale closures in the small-data regime
  for stable and accurate LES}. \textit{arXiv preprint arXiv:2201.07347}.

\bibitem[{Hairer and Mattingly(2006)Hairer, and
  Mattingly}]{hairer2006ergodicity}
Hairer, M., and J.~Mattingly, 2006: {Ergodicity of the 2D Navier-Stokes
  equations with degenerate stochastic forcing}. \textit{Annals of
  Mathematics}, 993--1032.

\bibitem[{Hairer and Mattingly(2011)Hairer, and Mattingly}]{hairer2011theory}
Hairer, M., and J.~Mattingly, 2011: {A theory of hypoellipticity and unique
  ergodicity for semilinear stochastic PDEs}. \textit{Electron. J. Probab.},
  658--738.

\bibitem[{Hornik et~al.(1989)Hornik, Stinchcombe,, and White}]{hornik1989}
Hornik, K., M.~Stinchcombe, and H.~White, 1989: Multilayer feedforward networks
  are universal approximators. \textit{Neural networks}, \textbf{2~(5)},
  359--366.

\bibitem[{Hussein et~al.(2017)Hussein, Gaber, Elyan,, and Jayne}]{hussein2017}
Hussein, A., M.~M. Gaber, E.~Elyan, and C.~Jayne, 2017: Imitation learning: A
  survey of learning methods. \textit{ACM Computing Surveys (CSUR)},
  \textbf{50~(2)}, 1--35.

\bibitem[{Ioffe and Szegedy(2015)Ioffe, and Szegedy}]{ioffe2015}
Ioffe, S., and C.~Szegedy, 2015: Batch normalization: Accelerating deep network
  training by reducing internal covariate shift. \textit{International
  conference on machine learning}, PMLR, 448--456.

\bibitem[{Jansen and Held(2014)Jansen, and Held}]{jansen2014parameterizing}
Jansen, M.~F., and I.~M. Held, 2014: Parameterizing subgrid-scale eddy effects
  using energetically consistent backscatter. \textit{Ocean Modelling},
  \textbf{80}, 36--48.

\bibitem[{Keisler(2022)}]{keisler2022}
Keisler, R., 2022: Forecasting global weather with graph neural networks.
  \textit{arXiv preprint arXiv:2202.07575}.

\bibitem[{Kochkov et~al.(2021)Kochkov, Smith, Alieva, Wang, Brenner,, and
  Hoyer}]{kochkov2021}
Kochkov, D., J.~A. Smith, A.~Alieva, Q.~Wang, M.~P. Brenner, and S.~Hoyer,
  2021: Machine learning--accelerated computational fluid dynamics.
  \textit{Proc. Natl. Acad. Sci. USA}, \textbf{118~(21)}, e2101784\,118.

\bibitem[{Kraichnan(1971)}]{kraichnan1971}
Kraichnan, R.~H., 1971: Inertial-range transfer in two-and three-dimensional
  turbulence. \textit{J.\ Fluid\ Mech.}, \textbf{47~(3)}, 525--535.

\bibitem[{Large et~al.(1994)Large, McWilliams,, and Doney}]{larg94}
Large, W., J.~C. McWilliams, and S.~C. Doney, 1994: Oceanic vertical mixing: a
  review and a model with a nonlocal boundary layer parameterization.
  \textit{{\it Rev. Geophys.}}, \textbf{32}, 363--403, {\footnotesize\tt
  doi:10.1029/94RG01872}.

\bibitem[{Lele(1992)}]{lele92}
Lele, S.~K., 1992: Compact finite difference schemes with spectral-like
  resolution. \textit{J.\ Comput. Phys.}, \textbf{103}, 16--42,
  {\footnotesize\tt doi:10.1016/0021-9991(92)90324-R}.

\bibitem[{List et~al.(2022)List, Chen,, and Thuerey}]{list2022}
List, B., L.-W. Chen, and N.~Thuerey, 2022: Learned turbulence modelling with
  differentiable fluid solvers: physics-based loss functions and optimisation
  horizons. \textit{Journal of Fluid Mechanics}, \textbf{949}, A25.

\bibitem[{Loshchilov and Hutter(2016)Loshchilov, and Hutter}]{loshchilov2016}
Loshchilov, I., and F.~Hutter, 2016: Sgdr: Stochastic gradient descent with
  warm restarts. \textit{arXiv preprint arXiv:1608.03983}.

\bibitem[{Ma et~al.(2019)Ma, Wang,, and Weinan}]{ma2019model}
Ma, C., J.~Wang, and E.~Weinan, 2019: Model reduction with memory and the
  machine learning of dynamical systems. \textit{Communications in
  Computational Physics}, \textbf{25~(4)}, 947--962.

\bibitem[{Ma et~al.(2018)Ma, Wang et~al.}]{ma2018}
Ma, C., J.~Wang, and Coauthors, 2018: Model reduction with memory and the
  machine learning of dynamical systems. \textit{arXiv preprint
  arXiv:1808.04258}.

\bibitem[{Marston et~al.(2008)Marston, Conover,, and Schneider}]{marston2008}
Marston, J., E.~Conover, and T.~Schneider, 2008: Statistics of an unstable
  barotropic jet from a cumulant expansion. \textit{Journal of the Atmospheric
  Sciences}, \textbf{65~(6)}, 1955--1966.

\bibitem[{Maulik et~al.(2019)Maulik, San, Rasheed,, and Vedula}]{maulik2019}
Maulik, R., O.~San, A.~Rasheed, and P.~Vedula, 2019: Subgrid modelling for
  two-dimensional turbulence using neural networks. \textit{Journal of Fluid
  Mechanics}, \textbf{858}, 122--144.

\bibitem[{McWilliams(1984)}]{mcwilliams1984}
McWilliams, J.~C., 1984: The emergence of isolated coherent vortices in
  turbulent flow. \textit{Journal of Fluid Mechanics}, \textbf{146}, 21--43.

\bibitem[{Miyanawala and Jaiman(2017)Miyanawala, and
  Jaiman}]{miyanawala2017efficient}
Miyanawala, T.~P., and R.~K. Jaiman, 2017: {An efficient deep learning
  technique for the Navier-Stokes equations: Application to unsteady wake flow
  dynamics}. \textit{arXiv preprint arXiv:1710.09099}.

\bibitem[{Orszag and Israeli(1974)Orszag, and Israeli}]{oris74}
Orszag, S.~A., and M.~Israeli, 1974: Numerical simulation of viscous
  incompressible flows. \textit{{\it Ann. Rev. fluid mechanics}}, \textbf{6},
  281--318, {\footnotesize\tt doi:10.1146/annurev.fl.06.010174.001433}.

\bibitem[{Ott et~al.(2020)Ott, Pritchard, Best, Linstead, Curcic,, and
  Baldi}]{ott2020}
Ott, J., M.~Pritchard, N.~Best, E.~Linstead, M.~Curcic, and P.~Baldi, 2020: A
  fortran-keras deep learning bridge for scientific computing.
  \textit{Scientific Programming}, \textbf{2020}.

\bibitem[{Parish and Duraisamy(2017)Parish, and Duraisamy}]{parish2017non}
Parish, E.~J., and K.~Duraisamy, 2017: {Non-Markovian closure models for large
  eddy simulations using the Mori-Zwanzig formalism}. \textit{Physical Review
  Fluids}, \textbf{2~(1)}, 014\,604.

\bibitem[{Pascal and Basdevant(1992)Pascal, and
  Basdevant}]{pascal1992nonlinear}
Pascal, F., and C.~Basdevant, 1992: Nonlinear {G}alerkin method and
  subgrid-scale model for two-dimensional turbulent flows. \textit{Theoretical
  and Computational Fluid Dynamics}, \textbf{3~(5)}, 267--284.

\bibitem[{Pearson et~al.(2017)Pearson, Fox-Kemper, Bachman,, and
  Bryan}]{pearson2017}
Pearson, B., B.~Fox-Kemper, S.~Bachman, and F.~Bryan, 2017: Evaluation of
  scale-aware subgrid mesoscale eddy models in a global eddy-rich model.
  \textit{Ocean Modelling}, \textbf{115}, 42--58.

\bibitem[{Perezhogin et~al.(2023)Perezhogin, Zanna,, and
  Fernandez-Granda}]{perezhogin2023}
Perezhogin, P., L.~Zanna, and C.~Fernandez-Granda, 2023: Generative data-driven
  approaches for stochastic subgrid parameterizations in an idealized ocean
  model. \textit{arXiv preprint arXiv:2302.07984}.

\bibitem[{Piomelli et~al.(1991)Piomelli, Cabot, Moin,, and
  Lee}]{piomelli1991subgrid}
Piomelli, U., W.~H. Cabot, P.~Moin, and S.~Lee, 1991: Subgrid-scale backscatter
  in turbulent and transitional flows. \textit{Physics of Fluids A: Fluid
  Dynamics}, \textbf{3~(7)}, 1766--1771.

\bibitem[{Rahaman et~al.(2019)Rahaman, Baratin, Arpit, Draxler, Lin, Hamprecht,
  Bengio,, and Courville}]{rahaman2019spectral}
Rahaman, N., A.~Baratin, D.~Arpit, F.~Draxler, M.~Lin, F.~Hamprecht, Y.~Bengio,
  and A.~Courville, 2019: On the spectral bias of neural networks.
  \textit{International Conference on Machine Learning}, PMLR, 5301--5310.

\bibitem[{Ramachandran et~al.(2017)Ramachandran, Zoph,, and
  Le}]{ramachandran2017searching}
Ramachandran, P., B.~Zoph, and Q.~V. Le, 2017: Searching for activation
  functions. \textit{arXiv preprint arXiv:1710.05941}.

\bibitem[{Rasp(2020)}]{rasp2020}
Rasp, S., 2020: Coupled online learning as a way to tackle instabilities and
  biases in neural network parameterizations: general algorithms and lorenz 96
  case study (v1. 0). \textit{Geoscientific Model Development},
  \textbf{13~(5)}, 2185--2196.

\bibitem[{Ross et~al.(2023)Ross, Li, Perezhogin, Fernandez-Granda,, and
  Zanna}]{ross2023}
Ross, A., Z.~Li, P.~Perezhogin, C.~Fernandez-Granda, and L.~Zanna, 2023:
  Benchmarking of machine learning ocean subgrid parameterizations in an
  idealized model. \textit{Journal of Advances in Modeling Earth Systems},
  \textbf{15~(1)}, e2022MS003\,258.

\bibitem[{Souza et~al.(2020)}]{souza2020}
Souza, A.~N., G.~Wagner, A.~Ramadhan, B.~Allen, V.~Churavy, J.~Schloss,
  J.~Campin, C.~Hill, A.~Edelman, J.~Marshall, and Coauthors, 2020: Uncertainty
  quantification of ocean parameterizations: Application to the
  k-profile-parameterization for penetrative convection. \textit{Journal of
  Advances in Modeling Earth Systems}, \textbf{12~(12)}, e2020MS002\,108.

\bibitem[{Srinivasan and Young(2012)Srinivasan, and Young}]{sryo2012}
Srinivasan, K., and W.~R. Young, 2012: Zonostrophic instability. \textit{J.\
  Atmos.\ Sci.}, \textbf{69}, 1633--1656, {\footnotesize\tt
  doi:10.1175/JAS-D-13-0246.1}.

\bibitem[{Srivastava et~al.(2014)Srivastava, Hinton, Krizhevsky, Sutskever,,
  and Salakhutdinov}]{srivastava2014}
Srivastava, N., G.~Hinton, A.~Krizhevsky, I.~Sutskever, and R.~Salakhutdinov,
  2014: Dropout: a simple way to prevent neural networks from overfitting.
  \textit{The journal of machine learning research}, \textbf{15~(1)},
  1929--1958.

\bibitem[{Stachenfeld et~al.(2021)}]{stachenfeld2021}
Stachenfeld, K., D.~B. Fielding, D.~Kochkov, M.~Cranmer, T.~Pfaff, J.~Godwin,
  C.~Cui, S.~Ho, P.~Battaglia, and A.~Sanchez-Gonzalez, 2021: Learned coarse
  models for efficient turbulence simulation. \textit{arXiv preprint
  arXiv:2112.15275}.

\bibitem[{Subel et~al.(2022)Subel, Guan, Chattopadhyay,, and
  Hassanzadeh}]{subel2022explaining}
Subel, A., Y.~Guan, A.~Chattopadhyay, and P.~Hassanzadeh, 2022: Explaining the
  physics of transfer learning a data-driven subgrid-scale closure to a
  different turbulent flow. \textit{arXiv preprint arXiv:2206.03198}.

\bibitem[{Xiao et~al.(2009)Xiao, Wan, Chen,, and Eyink}]{xiao2009}
Xiao, Z., M.~Wan, S.~Chen, and G.~Eyink, 2009: Physical mechanism of the
  inverse energy cascade of two-dimensional turbulence: a numerical
  investigation. \textit{Journal of Fluid Mechanics}, \textbf{619}, 1--44.

\bibitem[{Yuval and O'Gorman(2020)Yuval, and O'Gorman}]{yuval2020}
Yuval, J., and P.~A. O'Gorman, 2020: Stable machine-learning parameterization
  of subgrid processes for climate modeling at a range of resolutions.
  \textit{Nature communications}, \textbf{11~(1)}, 1--10.

\bibitem[{Yuval et~al.(2021)Yuval, O'Gorman,, and Hill}]{yuval2021}
Yuval, J., P.~A. O'Gorman, and C.~N. Hill, 2021: Use of neural networks for
  stable, accurate and physically consistent parameterization of subgrid
  atmospheric processes with good performance at reduced precision.
  \textit{Geophysical Research Letters}, \textbf{48~(6)}, e2020GL091\,363.

\bibitem[{Zanna and Bolton(2020)Zanna, and Bolton}]{zanna2020}
Zanna, L., and T.~Bolton, 2020: Data-driven equation discovery of ocean
  mesoscale closures. \textit{Geophysical Research Letters}, \textbf{47~(17)},
  e2020GL088\,376.

\bibitem[{Zhou(2020)}]{zhou2020}
Zhou, D.-X., 2020: Universality of deep convolutional neural networks.
  \textit{Applied and computational harmonic analysis}, \textbf{48~(2)},
  787--794.

\bibitem[{Zhou et~al.(2019)Zhou, He, Wang,, and Jin}]{zhou2019}
Zhou, Z., G.~He, S.~Wang, and G.~Jin, 2019: Subgrid-scale model for large-eddy
  simulation of isotropic turbulent flows using an artificial neural network.
  \textit{Computers \& Fluids}, \textbf{195}, 104\,319.

\end{thebibliography}

\end{document}